\documentclass[fleqn,usenatbib]{mnras}

\usepackage{newtxtext,newtxmath}

\usepackage[T1]{fontenc}

\DeclareRobustCommand{\VAN}[3]{#2}
\let\VANthebibliography\thebibliography
\def\thebibliography{\DeclareRobustCommand{\VAN}[3]{##3}\VANthebibliography}

\usepackage[normalem]{ulem}
\usepackage[dvipsnames]{xcolor}
\usepackage{graphicx}	
\usepackage{amsmath}	
\usepackage{amsmath,siunitx}
\usepackage{gensymb}
\usepackage{booktabs}   
\usepackage{multirow} 
\DeclareUnicodeCharacter{2212}{-}

\title[sGRB kilonova constraints] {A tale of two mergers: constraints on kilonova detection in two short GRBs at z$\sim$0.5
}

\author[O'Connor et al. (2020)]{
B. O'Connor$^{1,2,3,4}$\thanks{E-mail: oconnorb@gwu.edu}, E. Troja$^{3,4}$, S. Dichiara$^{3,4}$, E. A. Chase$^{5,6,7,8}$, G. Ryan$^{3,4}$, S. B. Cenko$^{4,9}$,   \newauthor
C. L. Fryer$^{5,10,11,12,13,1}$,  R. Ricci$^{14,15}$,  F. Marshall$^{4}$, C. Kouveliotou$^{1,2}$,  R. T. Wollaeger$^{5,11}$, \newauthor
C. J. Fontes$^{5,6}$, O. Korobkin$^{5,10,11}$, P. Gatkine$^{3,16}$,   A.  Kutyrev$^{3,4}$,  S. Veilleux$^{3,9}$, N. Kawai$^{17}$, \newauthor
  T. Sakamoto$^{18}$ \\
$^{1}$Department of Physics, The George Washington University, 725 21st Street NW, Washington, DC 20052, USA\\
$^{2}$Astronomy, Physics and Statistics Institute of Sciences (APSIS), The George Washington University, Washington, DC 20052, USA\\
$^{3}$Department of Astronomy, University of Maryland, College Park, MD 20742-4111, USA \\
$^{4}$Astrophysics Science 
Division, NASA Goddard Space Flight Center, 8800 Greenbelt Rd, Greenbelt, MD 20771, USA\\
$^{5}$Center for Theoretical Astrophysics, Los Alamos National Laboratory, Los Alamos, NM, 87545, USA \\
$^{6}$Computational Physics Division, Los Alamos National Laboratory, Los Alamos, NM, 87545, USA \\
$^{7}$Center for Interdisciplinary Exploration and Research in Astrophysics (CIERA), Northwestern University, Evanston, IL, 60201, USA \\
$^{8}$Department of Physics and Astronomy, Northwestern University, Evanston, IL, 60208, USA \\
$^{9}$Joint Space-Science Institute, University of Maryland, College Park, MD 20742 USA \\
$^{10}$Joint Institute for Nuclear Astrophysics, Center for the Evolution of the Elements, USA\\
$^{11}$Computer, Computational, and Statistical Sciences Division, Los Alamos National Laboratory, Los Alamos, NM, 87545, USA\\
$^{12}$Department of Astronomy, The University of Arizona, Tucson, AZ 85721, USA\\
$^{13}$Department of Physics and Astronomy, The University of New Mexico, Albuquerque, NM 87131, USA \\
$^{14}$Istituto Nazionale di Ricerche Metrologiche, 10135, Torino, Italy \\
$^{15}$INAF-Istituto di Radioastronomia, via Gobetti 101, 40129, Bologna, Italy \\
$^{16}$Department of Astronomy, California Institute of Technology, Pasadena, CA, USA \\
$^{17}$Department of Physics, Tokyo Institute of Technology,
2-12-1 (H-29) Ookayama, Meguro-ku, Tokyo 152-8551, Japan \\
$^{18}$Department of Physics and Mathematics, Aoyama Gakuin University, 5-10-1 Fuchinobe, Chuo-ku, Sagamihara-shi Kanagawa 252-5258, Japan \\
}

\date{Accepted XXX. Received YYY; in original form ZZZ}

\pubyear{2020}

\begin{document}
\label{firstpage}
\pagerange{\pageref{firstpage}--\pageref{lastpage}}
\maketitle

\begin{abstract}
We present a detailed multi-wavelength analysis of two short Gamma-Ray Bursts (sGRBs) detected by the \textit{Neil Gehrels Swift Observatory}: GRB 160624A at $z\!=\!0.483$ and GRB 200522A at $z\!=\!0.554$. 
These sGRBs demonstrate very different properties in their observed emission and environment. GRB~160624A is associated to a late-type galaxy with an old stellar population ($\approx$3 Gyr) and moderate on-going star 
formation ($\approx$1 $M_{\odot}$ yr$^{-1}$). 
\textit{Hubble} and Gemini limits on optical/nIR emission from GRB 160624A are among the most stringent for sGRBs, leading to tight constraints on the allowed kilonova properties. In particular, we rule out any kilonova brighter than AT2017gfo, disfavoring large masses of wind ejecta ($\lesssim$0.03 $M_\odot$). In contrast, observations of GRB 200522A uncovered a luminous ($L_\textrm{F125W}\approx 10^{42}$ erg s$^{-1}$ at 2.3~d) and red ($r-H\!\approx\! 1.3$ mag) counterpart. The red color can be explained either by bright kilonova emission powered by the radioactive decay of a large amount of wind ejecta (0.03 $M_\odot$ $\lesssim$ $M$ $\lesssim$ 0.1 $M_\odot$)
or moderate extinction, $E(B-V)\!\approx\!0.1-0.2$ mag, along the line of sight.
The location of this sGRB in the inner regions of a young ($\approx$0.1 Gyr) star-forming ($\approx$2-6 $M_{\odot}$  yr$^{-1}$) galaxy and the limited sampling of its counterpart do not allow us to rule out dust effects
as contributing, at least in part, to the red color. 
\end{abstract}



\clearpage

\section{Introduction}

The progenitors of short Gamma-Ray Bursts (sGRBs) were long suspected to be compact binary mergers \citep{Blinnikov1984,Paczynski1986,Eichler1989,Narayan1992}, comprising either two 
neutron stars (NSs; \citealt{Ruffert1999, Rosswog2003,Rosswog2005}) or a NS and a black hole (BH; \citealt{Faber2006,Shibata2011}). 
The merger remnant is either a BH \citep{Baiotti2008,Kiuchi2009} or a massive NS \citep{Giacomazzo2013,Hotokezaka2013nsremant}. In either case, the merger launches a relativistic jet which produces the observed prompt gamma-ray emission \citep{Rezzolla2011,Paschalidis2015,Ruiz2016}. The interaction of the relativistic jet with the surrounding medium produces the afterglow emission \citep{Meszaros1997,Sari1998,Wijers1999} observed across the electromagnetic (EM) spectrum. 

The connection between sGRBs and NS mergers was consolidated by the joint detection of the gravitational wave (GW)  event GW170817 \citep{Abbott2017} and the short GRB~170817A \citep{Goldstein2017, Savchenko2017}. These were followed by the luminous 
($L_\textrm{bol}\,\approx\!10^{42}$ erg s$^{-1}$) kilonova AT2017gfo \citep{Andreoni2017,Arcavi2017,Chornock2017,Coulter2017,Covino2017,Cowperthwaite2017,Drout2017,Evans2017, Kasliwal2017, Lipunov2017,Nicholl2017,Pian2017,Shappee2017,Smartt2017,Tanvir2017,Troja2017,Utsumi2017,Valenti2017}. 
AT2017gfo was initially  characterized by a blue thermal spectrum, which progressively shifted to redder colors and displayed broad undulations typical of fast moving ejecta \citep[e.g.,][]{Watson2019}. 

Kilonova emission following a NS-NS merger originates from the radioactive decay of freshly synthesized r-process elements in  neutron-rich matter surrounding the remnant compact object \citep{Li1998,Metzger2010,Barnes2013,Tanaka2013,Grossman2014,Kasen2017}.
Kilonovae are hallmarked by ``blue'' thermal emission within a day of merger (e.g., AT2017gfo) which fades and gives way to the ``red'' and near-infrared (nIR) emission, persisting for roughly a week post-merger. Neutron-rich material (electron fraction $Y_e\!<\!0.25$) composed of high-opacity lanthanides produces the red component, while the blue component results from ejecta material with higher electron fractions \citep{Barnes2013,Kasen2015,Kasen2017,Tanaka2017,Wollaeger2018,Wollaeger2019,Fontes2020,Even2020,Korobkin2020}. Dynamical ejecta, tidally stripped from the approaching neutron star(s), primarily contributes to the red component. In addition, a portion of the matter that congregates in an accretion disk surrounding the remnant compact object is released as wind ejecta \citep{Metzger2008,Dessart2009,Lee2009,Fernandez2013,Perego2014,Just2015,miller2019}.
Ejecta in the disk supports a wide range of electron fractions, enhancing either a blue or red kilonova. The identity of the merger remnant influences the disk, with a longer-lived high mass neutron star remnant increasing the electron fraction of disk ejecta \citep{Kasen2015}. The range of electron fractions of ejecta predicted from models of disk winds varies with the implementation of the neutrino transport \citep{miller2019}. 

Although kilonovae are usually described as near-isotropic, they exhibit viewing-angle dependent variations based on ejecta morphology \citep{Korobkin2020} and lanthanide curtaining effects \citep{Kasen2015}.
Observations of AT2017gfo were possible thanks to the particular geometry of GW170817, whose relativistic jet was misaligned 
with respect to our line of sight \citep{Troja2017, Lazzati2017,Mooley2018,Troja19,Ghirlanda2019,Lamb2019gwevent}. As a consequence, its afterglow appeared at later times and remained relatively dim, allowing for a complete view of the kilonova. 
Had the same event been seen closer to its jet's axis (on-axis), the GRB afterglow would have outshined any kilonova emission. 

The majority of sGRBs are discovered at much larger distances than GW170817 and are observed close to their jet's axis \citep{Ryan15,Beniamini2019,BeniaminiNakar2019,Wu2019}. Their bright broadband afterglow is often the dominant emission component, which complicates the identification of any associated kilonova \citep[see, e.g.,][]{Yang2015,Jin2015,Ascenzi2019kn}.
Whereas the range of luminosities and timescales of kilonova emission largely overlaps with standard GRB afterglows, the red color of a kilonova \citep{Barnes2013,Tanaka2013}, much redder than any typical afterglow, is one of its distinctive features.
Nonetheless, even the color information may be insufficient  for an unambiguous identification.  A counterpart with unusually red colors was found for the short GRB~070724A \citep[$i$-$K_s$$\approx$4;][]{Berger2008} and GRB~071227 \citep[$r$-$z$$\approx$1.5;][]{Eyles2019} and,
in both cases, attributed to dust effects at the GRB site. 
The rapid timescales and high luminosity ($\approx\! 10^{43}$ erg s$^{-1}$) of these two sources did not match the predictions of a radioactive-powered kilonova, although they could fit within the expected range for a magnetar-powered kilonova \citep{Yu2013}. 

Densely sampled multi-color observations, extending to the nIR range, 
proved to be essential in the identification of the kilonova candidates 
GRB~130603B \citep{Tanvir2013} and GRB~160821B \citep{Troja2019,Lamb2019}. The counterpart of GRB~130603B was identified within the spiral
arm of its bright host galaxy. The source appeared unusually red, 
in part due to significant presence of dust along the sightline ($A_V\!\sim$1 mag; \citealt{deUgartePostigo2014}), and was seen to evolve over the course of time, from 
$R$-$H\approx$1.7 $\pm$ 0.15 at about 14 hr to $R$-$H$ $>$ 2.5 at about 9 d. 
GRB~160821B was instead located in the outskirts of a nearby spiral galaxy, and its counterpart was also identified as unusually red \citep[$V$–$K$\!$\approx$1.9;][]{Kasliwal2017a}. 
A detailed modeling of the X-ray and radio afterglow was able to disentangle the presence of an additional emission component in the optical and nIR data, slightly less luminous than AT2017gfo and with similar timescales and color evolution \citep{Troja2019,Lamb2019}. 
For both GRB~130603B and GRB~160821B, a good spectral sampling over multiple epochs was a fundamental ingredient to distinguish the kilonova candidate from the underlying bright afterglow. 

In addition to these candidate kilonovae, there are a a number of claimed kilonova detections based on an optical excess, e.g., GRB 050709 \citep{Jin2016}, GRB 060614 \citep{Yang2015}, GRB 070809 \citep{Jin2020}, GRB 080503 \citep{Perley2009}, and GRB 150101B \citep{Troja150101B}. The situation for these events is less clear due to the lack of deep nIR observations, critical to distinguish kilonova emission from standard afterglow.

The number of sGRBs with well characterized afterglows and sensitive nIR observations is still restricted to a handful of cases 
\citep[see, e.g.,][]{Gompertz2018,Rossi2020}.
In this work, we continue filling this observational gap by presenting a detailed multi-wavelength study of two distant sGRBs: GRB 160624A at $z\!=\!0.483$ and GRB 200522A at $z\!=\!0.554$. 
We complement the early {\it Swift} data with deep \textit{Chandra} observations in order to characterize the afterglow temporal evolution up to late times, and use deep Gemini and \textit{HST} imaging to search for kilonova emission. 

The paper is organized as follows. In \S \ref{sec: analysis_both}, we present the observations and data analysis for GRBs 160624A and 200522A. In \S \ref{sec: methods}, we describe the methods applied for our afterglow and kilonova modeling, as well as the galaxy spectral energy distribution (SED) fitting procedure. The results are presented in \S \ref{sec: results}, and our conclusions in \S \ref{sec: conclusions}. For each GRB, we provide the time of observations relative to the BAT trigger time, $T_0$, in the observer frame. All magnitudes are presented in the AB system. We adopt the standard $\Lambda$-CDM cosmology with parameters $H_0=67.4$, $\Omega_M=0.315$, and $\Omega_\Lambda=0.685$ \citep{Planck2018}. All confidence intervals are at the $1\sigma$ level and upper limits at the $3\sigma$ level, unless otherwise stated. Throughout the paper we adopt the convention $F_\nu\propto t^{-\alpha}\nu^{-\beta}$.

\section{Observations and Analysis}
\label{sec: analysis_both}
\subsection{GRB 160624A}
\subsubsection{Gamma-ray Observations}
\begin{figure} 
\centering
\includegraphics[width=\columnwidth]{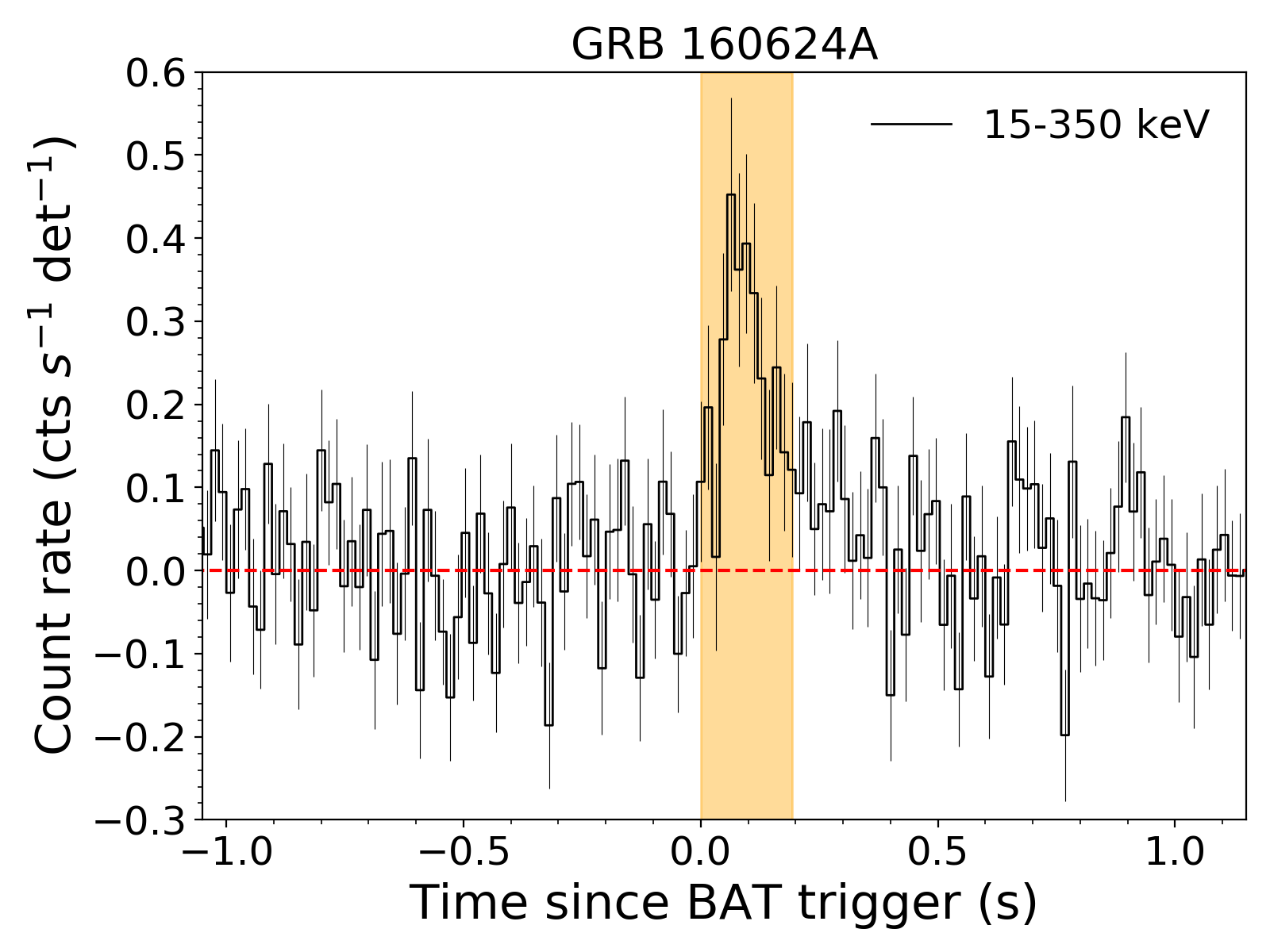}
\caption{BAT mask-weighted lightcurve (15-350 keV) of GRB 160624A with 16 ms binning. The shaded vertical region marks the $T_{90}$ duration.}
\label{fig: 160624A_bat}
\end{figure}

GRB 160624A triggered the \textit{Swift} Burst Alert Telescope \citep[BAT;][]{Barthelmy2005} at 2016 June 24 11:27:01 UT \citep{Dai2016}, hereafter referred to as $T_0$ for this GRB. The burst was single pulsed with duration $T_{90}= 0.2 \pm 0.1$~s and fluence $f_\gamma=(4.0\pm 0.9)\times 10^{-8}$ erg cm$^{-2}$ (15-150 keV). We performed a search of the BAT lightcurve (Figure \ref{fig: 160624A_bat}) for extended emission \citep[EE;][]{Norris2006} following the prompt phase. The search yields no evidence for EE with an upper limit of $<2.3\times10^{-7}$ erg cm$^{-2}$ (15-150 keV) from $T_0+2$ s and $T_0+100$~s. 

GRB 160624A was also detected by the \textit{Fermi} Gamma-ray Burst Monitor \citep[GBM;][]{Meegan2009}. The time-averaged GBM spectrum, from $T_0-0.06$ s to $T_0+0.2$, is well fit by a power-law with an exponential cutoff with low-energy spectral index $\alpha=-0.4 \pm 0.3$ and a peak energy $E_p=800\pm 400$ keV \citep{Hamburg2016}. 
Based on this model, the observed fluence is $f_\gamma =(5.2\pm0.5)\times 10^{-7}$ erg cm$^{-2}$ (10-1,000 keV), 
corresponding to an isotropic equivalent gamma-ray energy $E_{\gamma,\textrm{iso}}=(4.7\pm1.5)\times 10^{50}$ erg (1 keV - 10 MeV; rest frame) at a redshift $z\!=\!0.483$ (see \S \ref{sec: 160624A spectrum} and \ref{sec: 160624A_host}).

\begin{table*}
 	\centering
 	\caption{Log of Radio, Optical, nIR, and X-ray Observations of GRB 160624A. Upper limits correspond to a $3\sigma$ confidence level. 
 	}
 	\label{tab: observations}
 	\begin{tabular}{lccccccc}
    \hline
    \hline
   \multicolumn{8}{c}{\textbf{X-ray Observations}}  \\
   \hline
   \textbf{Start Date (UT)}  & \textbf{$\delta T$ (d)} & \textbf{Telescope} & \textbf{Instrument} &  & \textbf{Exposure (s)} & \textbf{Flux ($10^{-11}$ erg cm$^{-2}$ s$^{-1}$)} &\textbf{Flux Density$^{a}$ ($\mu$Jy)}\\
 \hline
 2016-06-24 11:28:00  & 0.00068 & \textit{Swift} & XRT &   & 0.6& $110\pm40$ & $110\pm40$ \\
2016-06-24  11:28:00 &0.00069 & \textit{Swift} & XRT & & 1.2  & $60\pm18$ & $60\pm18$\\
 2016-06-24 11:28:02  &0.00071 & \textit{Swift} & XRT &    & 1.3   & $56\pm18$  & $56\pm18$ \\
 2016-06-24 11:28:04  &0.00073 & \textit{Swift} & XRT &  & 2.4  &  $30\pm9$  & $30\pm9$\\
 2016-06-24 11:28:06  & 0.00075& \textit{Swift} & XRT & & 1.7  & $43\pm14$ &  $43\pm14$\\
 2016-06-24  11:28:08 &0.00077 & \textit{Swift} & XRT &  &  2.0 & $57\pm14$ &  $57\pm14$\\
 2016-06-24 11:28:21  &0.00093 & \textit{Swift} & XRT &  & 5.5  & $40\pm8$  & $40\pm8$ \\
 2016-06-24  11:28:26 & 0.00099& \textit{Swift} & XRT &   & 4.4  & $50\pm10$  & $50\pm10$ \\
 2016-06-24 11:28:32  & 0.00105& \textit{Swift} & XRT &  &  6.5 & $35\pm6$  & $35\pm6$\\
 2016-06-24 11:28:38  & 0.00112& \textit{Swift} & XRT &    &  6.7 & $32\pm6$ & $32\pm6$ \\
 2016-06-24 11:28:44 &0.00119 & \textit{Swift} & XRT & & 4.9  & $47\pm9$  & $47\pm9$\\
     2016-06-24 11:28:48& 0.00124& \textit{Swift} & XRT &  & 4.1  & $55\pm10$  & $55\pm10$ \\
 2016-06-24  11:28:54 &0.00131 & \textit{Swift} & XRT &   & 7.7  & $28\pm5$  & $28\pm5$ \\
 2016-06-24 11:29:01  &0.00139 & \textit{Swift} & XRT &  &  5.4 & $42\pm8$  & $42\pm8$ \\
 2016-06-24  11:29:08 &0.00146 & \textit{Swift} & XRT &    & 7.4  &$30
 \pm6$  & $30
 \pm6$ \\
 2016-06-24 11:29:14  & 0.0015 & \textit{Swift} & XRT & &  6.6 & $34\pm6$ & $34\pm6$\\
 2016-06-24  11:29:23 &0.0017  & \textit{Swift} & XRT &  & 10  & $22\pm4$  & $22\pm4$\\
 2016-06-24  11:29:42 & 0.0019& \textit{Swift} & XRT &   & 15   &$16\pm4$  & $16\pm4$ \\
 2016-06-24 11:29:56  &0.0020 & \textit{Swift} & XRT &   &  15 & $15\pm4$  & $15\pm4$\\
 2016-06-24   11:30:18   &0.0023 & \textit{Swift} & XRT &  &  22 & $9.4\pm2.5$  & $9.4\pm2.5$ \\
 2016-06-24 11:30:55  & 0.0025 & \textit{Swift} & XRT &   &   28 &$7.7\pm2.0$   & $7.7\pm2.0$\\
 2016-06-24 11:31:6  & 0.0028& \textit{Swift} & XRT &   & 30  &  $4.0\pm1.0$  & $4.0\pm1.0$ \\
  2016-06-24  11:31:55    & 0.0034 & \textit{Swift} & XRT &  & 85  &$1.4\pm0.4$  & $1.4\pm0.4$\\
 2016-06-24  11:33:23 & 0.0044 & \textit{Swift} & XRT &    & 176  & $0.26^{+0.13}_{-0.10}$   & $0.26^{+0.13}_{-0.10}$ \\
 2016-06-24  12:59:05 & 0.064 & \textit{Swift} & XRT &  & 2209  & $<0.05$  & $<0.05$ \\
  2016-06-24  14:36:44 &0.13 & \textit{Swift} & XRT &     & 2472  & $<0.04$  & $<0.04$ \\
 2016-06-24 23:58:38  &0.52 & \textit{Swift} & XRT & & 1113  &  $<0.07$  & $<0.07$ \\
 2016-06-25 0:21:17  & 0.54 & \textit{Swift} & XRT &   & 1227  & $<0.06$  & $<0.06$\\
 2016-06-25 01:57:21   & 0.60 & \textit{Swift} & XRT &   & 2458  & $<0.03$ & $<0.03$\\
  2016-07-02 22:59:34& 8.48 & \textit{Chandra} & ACIS-S3 &    & 47000  & $<2.3\times10^{-4}$  & $<2.3\times10^{-4}$ 
  \\
 \hline
 \hline
    \multicolumn{8}{c}{\textbf{Optical/nIR Photometry}}  \\
       \hline
   \textbf{Start Date (UT)}  & \textbf{$\delta T$ (d)} & \textbf{Telescope} & \textbf{Instrument} & \textbf{Filter} & \textbf{Exposure (s)} & \textbf{AB Mag} &\textbf{Flux Density$^{a}$ ($\mu$Jy)}\\
    \hline
    2016-06-24  11:28:18 & 0.002 & \textit{Swift} & UVOT & \textit{wh} & 147  & >21.8 & $<9.4$\\
    2016-06-24  11:31:49   & 0.004 & \textit{Swift} & UVOT & \textit{u} & 234 & >21.0 & $<17.7$\\
  2016-06-24 11:58:22 & 0.022 & Gemini-N & GMOS  & \textit{r} & 180 & >25.1 & $<0.38$\\
 2016-06-24 12:30:26 & 0.044 & Gemini-N & GMOS  & \textit{r} & 900 & >25.9 & $<0.18$ \\
2016-06-25 13:38:42 & 1.09 & Gemini-N & GMOS & \textit{r} & 1440 & >25.8 & $<0.20 $\\
 2016-06-27 19:35:49 & 3.34 &  \textit{HST} & ACS/WFC & \textit{F606W} & 1960  &>27.5 & $<0.04$ \\
 2016-06-28 17:49:54 & 4.27 & \textit{HST}  & WFC3  & \textit{F125W}   & 2411   & >27.2 & $<0.05$ \\
 2016-06-28 19:25:15 & 4.33 & \textit{HST} & WFC3 & \textit{F160W} & 2411  & >27.0 & $<0.06$\\
 2016-07-02 15:39:35 & 8.18 & \textit{HST}  & ACS/WFC & \textit{F606W} & 1960  & >27.5 & $<0.04$ \\
 2016-07-02 18:49:16 & 8.31 &\textit{HST} & WFC3 & \textit{F125W}   & 2411 &>27.2 & $<0.05$ \\
 2016-07-03 20:15:09 & 9.37 & \textit{HST}  & WFC3 & \textit{F160W}   & 2411  & >27.0 & $<0.06$ \\
 \hline
 \hline
    \multicolumn{8}{c}{\textbf{Optical Spectroscopy}}  \\
       \hline
   \textbf{Start Date (UT)}  & \textbf{$\delta T$ (d)} & \textbf{Instrument} & \textbf{Grating} & \textbf{$\lambda_\textrm{cen}$}   &   \textbf{Exposure (s)} & \textbf{Slit Width} &\\
   \hline
   2016-06-24 12:12:58 & 0.032 & GMOS & R400  & 600 nm   &    900 & $1\arcsec$ & \\
      \hline
      \hline
   \multicolumn{8}{c}{\textbf{Radio Observations}}  \\
      \hline
   \textbf{Start Date (UT)}  & \textbf{$\delta T$ (d)} & \textbf{Telescope} & \textbf{Configuration}  & \textbf{Filter} & \textbf{Exposure (s)} &  &\textbf{Flux Density$^{a}$ ($\mu$Jy)}\\
    \hline
     2016-06-25 11:47:08 & 1.01 & VLA & B & X (10 GHz) &  2820 &  & $<18$ \\
     \hline
    \end{tabular}
\begin{flushleft}
    \quad \footnotesize{$^a$ Optical/nIR fluxes were corrected for Galactic extinction due to interstellar reddening $E(B-V)=0.06$ mag \citep{Schlafly2011} using the extinction law by \citet{Fitzpatrick1999,Indebetouw2005}. X-ray fluxes were corrected for Galactic absorption $N_H=9.14\times 10^{20} $ cm$^{−2}$ \citep{Willingale2013}, and converted into flux densities at 1 keV using photon index $\Gamma=1.76$. } 
\end{flushleft}
\end{table*}

\begin{table}
 	\centering
 	\caption{Observations of the host galaxy of GRB 160624A. Magnitudes are not corrected for Galactic extinction, $E(B-V)=0.06$ mag, in the direction of the burst. 
 	}
 	\label{tab: observations_host}
 	\begin{tabular}{lccc}
    \hline
    \hline
 \textbf{Instrument} & \textbf{Filter} & \textbf{Exp. (s)} & \textbf{AB Mag} \\
\hline
  LDT/LMI & \textit{g} & 180 & $23.31\pm0.09$  \\
 \textit{HST}/ACS/WFC & \textit{F606W} & 1960  & $22.147\pm0.014$  \\
 Gemini/GMOS & \textit{r} & 1440 & $22.18\pm0.02$ \\ 
 LDT/LMI & \textit{r} & 180  & $22.16\pm0.08$ \\
  LDT/LMI & \textit{i}  & 180  & $21.66\pm0.08$  \\
 LDT/LMI & \textit{z}  & 360  & $21.47\pm0.08$ \\
 Gemini/NIRI & \textit{Y}& 540 & $21.42 \pm 0.15$ \\
\textit{HST}/WFC3 & \textit{F125W}   & 2411 & $20.842\pm0.004$ \\
 \textit{HST}/WFC3 & \textit{F160W}   & 2411  & $20.566\pm0.004$  \\
 Gemini/NIRI & \textit{$K_s$} & 180 & $20.32 \pm 0.08 $ \\
     \hline
    \end{tabular}
\end{table}

\begin{figure*} 
\centering
\includegraphics[width=1.7\columnwidth]{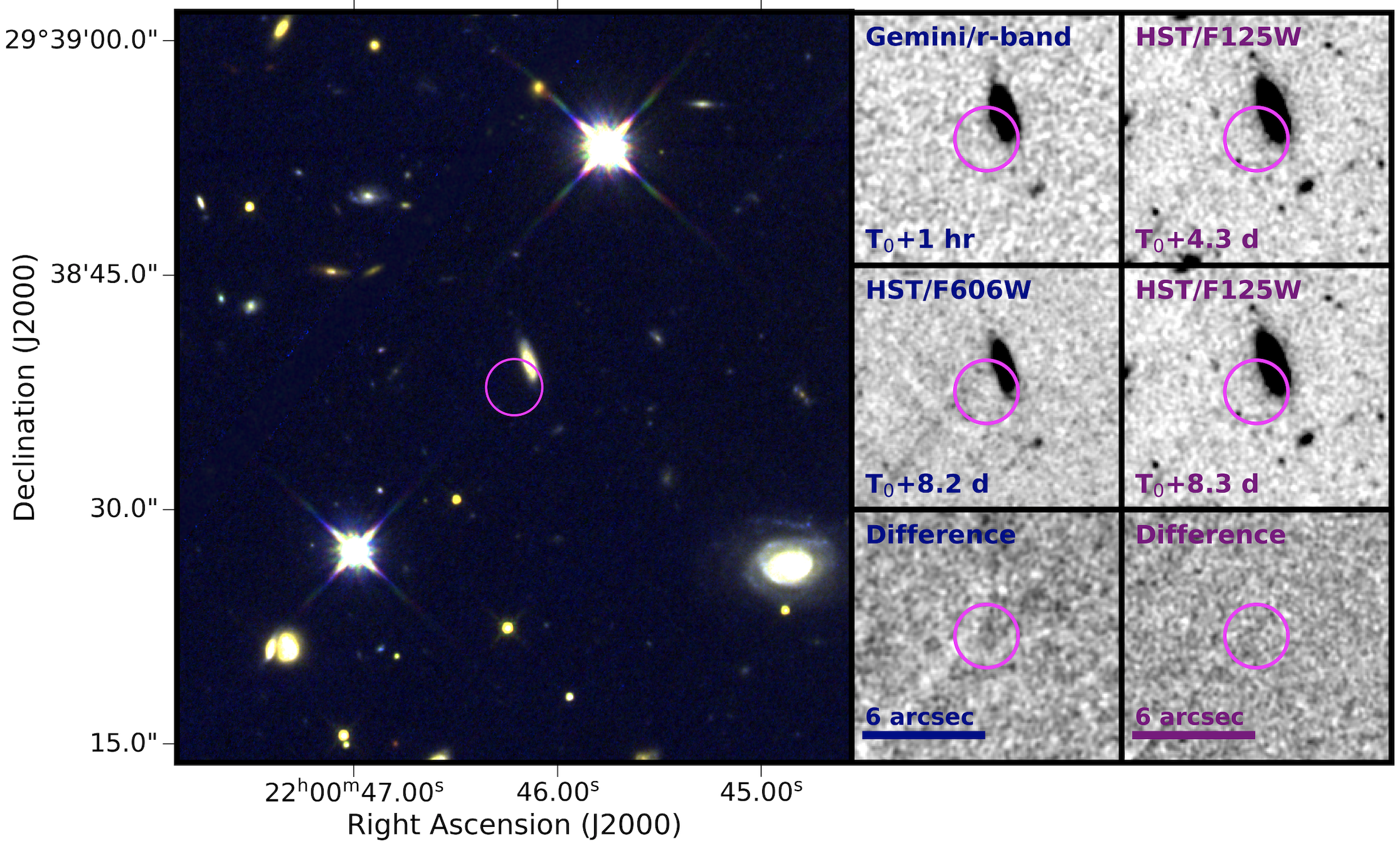}
\caption{\textbf{Left:} RGB image of the field of GRB 160624A using the three \textit{HST} filters: $F606W$ = blue, $F125W$ = green, and $F160W$= red. The XRT localization ($1.7\arcsec$) is shown in magenta. In the top left a chip gap from the F6060W observation is marginally visible.
\textbf{Center:} Deep images of the position of GRB 160624A in the Gemini/\textit{r-}band at $T_0+1$ hour (top) and \textit{HST}/$F606W$ filter at $T_0+8.2$ d (center). The bottom panel shows the difference image between Gemini and \textit{HST} (template) using \texttt{HOTPANTS}. There are no significant residuals within the enhanced XRT position. The images are smoothed for display purposes.
\textbf{Right:} \textit{HST} images of GRB 160624A in the $F125W$ filter taken at $T_0+4.3$ d (top) and $T_0+8.3$ d (center). The difference image is shown in the bottom panel. 
}
\label{fig: 160624A_RGB}
\end{figure*}

\subsubsection{X-ray Observations}
\label{sec: 160624A_X-ray_obs}

The \textit{Swift} X-ray Telescope \citep[XRT;][]{Burrows2005} began observing at $T_0+59$~s and localized
the X-ray afterglow at a position\footnote{https://www.swift.ac.uk/xrt\_curves/} RA, DEC (J2000) = $22^{h}00^m 46^{s}.21$, $+\ang{29;38;37.8}$ with an accuracy of \ang{;;1.7} (90\% confidence level (CL); \citealt{Evans2007,Evans2009}). Data were collected in Window Timing (WT) mode during the first 150~s 
and, as the source rapidly decreased in brightness, in Photon Counting (PC) mode. 
A deeper observation was carried out at $T_0$ + 8.5 d (PI: Troja; ObsId 18021) by the  \textit{Chandra} X-ray Observatory (ACIS-S3), but no X-ray counterpart was detected. 
We describe the observed temporal decay (Figure \ref{fig: 160624A_lightcurve}) with a broken power-law,
consisting of two segments with $F_x\propto t^{-\alpha_i}$. The initial decay is $\alpha_1=0.6\pm0.3$ which steepens to $\alpha_2=4.0\pm0.3$ after $t_\textrm{break}\sim 140$ s.

We model the XRT spectra with \texttt{XSPEC v12.10.1} \citep{Arnaud1996} by minimizing the Cash statistics \citep{Cash1979}. 
The Galactic hydrogen column density was fixed to the value $N_\textrm{H}=9.14\times10^{20}$ cm$^{-2}$ \citep{Willingale2013}. We determine that the time-averaged X-ray spectrum is well described (C-stat=249 for 336 dof) by an absorbed power-law model with photon index $\Gamma=\beta+1=1.76\pm0.15$ and intrinsic hydrogen column density $N_{H,\textrm{int}}=(2.8_{-0.6}^{+0.7})\times 10^{21}$ cm$^{-2}$ (required at the $5\sigma$ level). This yields a time-averaged unabsorbed flux $F_X=(3.5\pm0.3)\times 10^{-10}$ for WT mode data ($T_0+58$ to $T_0+150$ s) and $(2.1\pm0.2)\times 10^{-12} $ erg cm$^{-2}$ s$^{-1}$ (0.3-10 keV) for PC mode data ($T_0+150$ to $T_0+600$ s). The unabsorbed energy conversion factor (ECF) is $6.7\times10^{-11}$ erg cm$^{-2}$ cts$^{-1}$  for WT mode data and $6.9\times10^{-11}$ erg cm$^{-2}$ cts$^{-1}$ for PC mode. 

The \textit{Chandra} data were re-processed using the \texttt{CIAO} 4.12 data reduction package with \texttt{CALDB} Version 4.9.0, and filtered to the energy range $0.5-7$ keV. We corrected the native \textit{Chandra} astrometry by aligning the image with the \textit{Gaia} Data Release 2 \citep{GaiaDR2}. We utilized \texttt{CIAO} tools to extract a count-rate within the XRT error region (\ang{;;1.7} source aperture radius), utilizing nearby source-free regions to estimate the background. We detect zero counts in the source region with an estimated background of 0.6 counts yielding a $3\sigma$ upper limit of $1.2\times 10^{-4}$ cts s$^{-1}$ \citep{Kraft1991}. We convert this rate to $1.9\times10^{-15}$ erg cm$^{-2}$ s$^{-1}$ in the 0.3-10 keV band using the best fit spectral parameters. The derived X-ray fluxes are reported in Table \ref{tab: observations}.

\subsubsection{Optical/nIR Imaging} 
\label{sec: 160624A_optical_analysis}

The Ultra-Violet Optical Telescope \citep[UVOT;][]{Roming2005} on-board \textit{Swift} began observations at $T_0+77$ s although no optical afterglow is identified within the XRT position to $wh\geq 21.8$ AB mag \citep{dePasquale2016}.
The field was imaged with the Gemini Multi-Object Spectrograph \citep[GMOS;][]{Hook2004}
on the 8.1-meter Gemini North telescope (PI: Cucchiara) starting at $T_0+31$ min. 
An initial 180 second \textit{r}-band exposure led to the identification of a candidate host galaxy within the XRT error region \citep[SDSS J220046.14+293839.3;][]{Cucchiara2016}; for further discussion of the host association see \S \ref{sec: 160624A_host}. This was followed by deeper observations (900~s and 1440~s, respectively) at $T_0+1$~hr and $T_0+1$~d. Seeing during the observations was $\sim 0.5\arcsec$ with mean airmass 1.1 and 1.0, respectively. 
We retrieved the data from the Gemini archive. Data were analyzed following standard CCD reduction techniques and using the \texttt{Gemini} IRAF\footnote{IRAF is distributed by the National Optical Astronomy Observatory, which is operated by the Association of Universities for Research in Astronomy (AURA) under cooperative agreement with the National Science Foundation (NSF).} reduction package.

At later times, we performed two epochs of observations (PI: Troja; ObsId 14357) with the \textit{Hubble Space Telescope} (\textit{HST}) Advanced Camera for Surveys (ACS) Wide Field Camera (WFC) and Wide Field Camera 3 (WFC3) in Infrared (IR). See Table \ref{tab: observations} for a log of observations.
 The \textit{HST} data were processed using the \texttt{sndrizpipe}\footnote{https://github.com/srodney/sndrizpipe} pipeline, which makes use of standard procedures within the \texttt{DrizzlePac} package, in order to align, drizzle, and combine exposures.  The final image pixel scale was \ang{;;0.09}/pix for WFC3 (i.e., $F125W$ and $F160W$) and \ang{;;0.04}/pix for ACS (i.e., $F606W$). 

We identify no candidate counterpart within the XRT localization in either Gemini or \textit{HST} images. 
Since the XRT localization overlaps significantly with the candidate host galaxy (see Figure \ref{fig: 160624A_RGB}), we performed image subtraction between epochs using the High Order Transform of Psf ANd Template Subtraction code \citep[\texttt{HOTPANTS};][]{Becker2015} to search for transient sources embedded within the host galaxy's light. 
Due to the short time delay between Gemini epochs, our analysis
may not reveal a slowly evolving transient. We therefore 
verified our results using the late ($T_0+8.2$ d) \textit{HST}/$F606W$ image as the template. Furthermore, as kilonovae can dominate at either early or late times, depending on the composition of the ejecta, we performed image subtraction between the \textit{HST} epochs using each epoch as a template image.
No significant residual source was uncovered in either the Gemini or \textit{HST} difference images at any epoch, as shown in Figure \ref{fig: 160624A_RGB}. 

In order to determine the upper limit on a transient source in these images, we injected artificial point-like sources within the XRT position
and performed image subtraction to detect any residual signal. 
Gemini magnitudes were calibrated to nearby Sloan Digital Sky Survey Data Release 12 \citep[SDSS;][]{Fukugita1996} stars. 
\textit{HST} magnitude zeropoints were determined with the photometry keywords obtained from the \textit{HST} image headers, and were corrected with the STScI tabulated encircled energy fractions. The upper limits derived for the field are presented in Table \ref{tab: observations}. Upper limits within the galaxy's light are shallower by 0.3-0.5 mag.

Finally, we obtained imaging of the candidate host galaxy on May 20, 2018 ($T_0+1059$ d) with the Large
Monolithic Imager (LMI) mounted on the 4.3-meter Lowell Discovery Telescope (LDT) in Happy Jack, AZ. Observations were taken in the \textit{griz} filters, with seeing $\sim1.65\arcsec$ at a mean airmass of $\sim1.3$. We applied standard procedures for reduction and calibration of these images. We obtain the galaxy apparent magnitude in each filter using the \texttt{SExtractor} \texttt{MAG\_AUTO} parameter, which utilizes Kron elliptical apertures \citep{Bertin1996}. Magnitudes were calibrated to nearby SDSS stars, and reported in Table \ref{tab: observations_host}.
Near-infrared imaging  in the \textit{Y}\textit{$K_s$}~bands was carried out on July 25, 2020 with the Near-Infrared Imager \citep[NIRI;][]{Hodapp2003} on the 8-m Gemini North telescope. Data were reduced using standard procedures within the \texttt{DRAGONS} package. The photometry was calibrated to nearby sources from the Panoramic Survey Telescope and Rapid Response
System \citep[Pan-STARRS;][]{Chambers2016} and Two Micron All Sky Survey \citep[2MASS;][]{Skrutskie2006} for the \textit{Y-} and \textit{$K_s$-}band images, respectively. We used the offsets from \citet{Blanton2007} to convert 2MASS Vega magnitudes to the AB system.

\begin{figure} 
\centering
\includegraphics[width=\columnwidth]{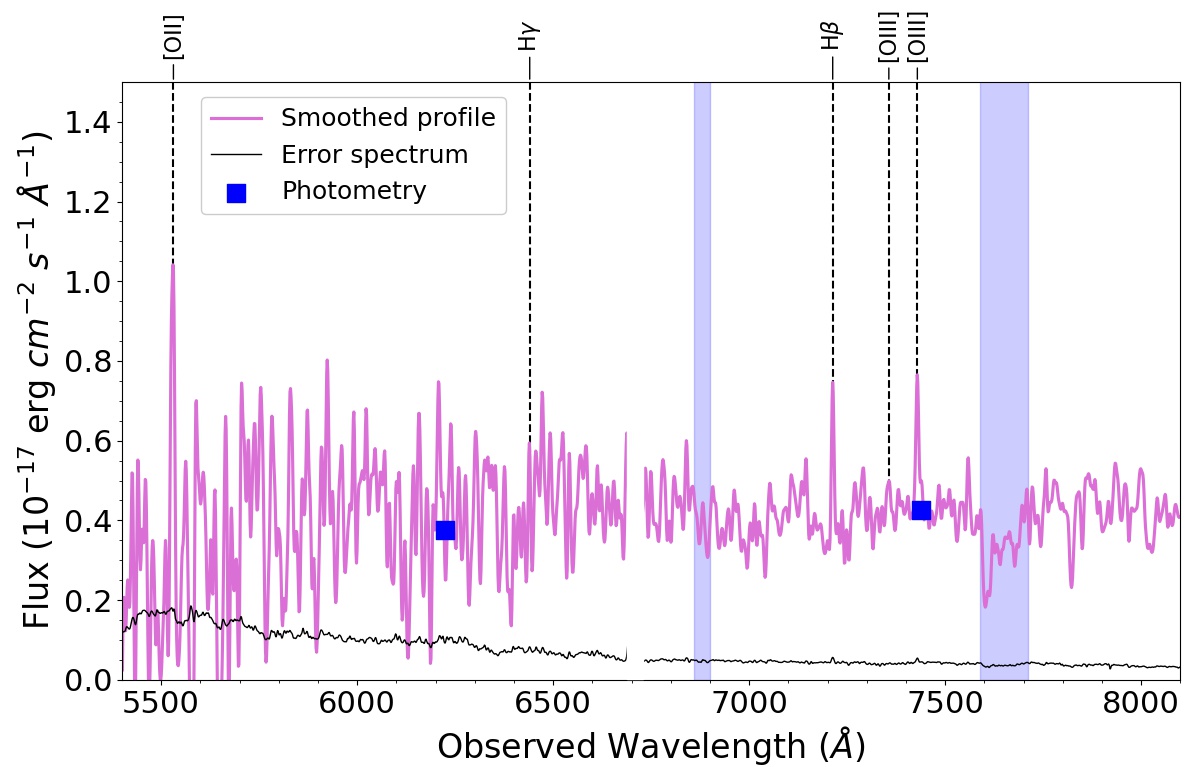}
\caption{Gemini North GMOS spectrum of the host galaxy of GRB 160624A at $z=0.4833\pm0.0004$. The spectrum has been smoothed with a Gaussian kernel, presented in purple, and the error spectrum is shown in black. Gemini/\textit{r}-band and LDT/\textit{i}-band photometry are shown in blue. Telluric absorption regions are marked by blue bands, and the blank region at $\sim 6710\,\si{\angstrom}$ is a chip gap. The positions of detected emission lines are indicated by dashed black lines. We do not detect the [OIII]$_{\lambda4959}$ line, but demonstrate its location for completeness.
}
\label{fig: 160624A_host_spectrum}
\end{figure}

\subsubsection{Optical Spectroscopy}
\label{sec: 160624A spectrum}

A spectrum of the candidate host galaxy was obtained using Gemini/GMOS (PI: Cucchiara)
starting at $T_0+46$ min. GMOS was configured with the R400 grating at a central wavelength of 600 nm. 
We reduced and analyzed the data using the Gemini IRAF package (v. 1.14). The resulting spectrum is shown in Figure \ref{fig: 160624A_host_spectrum}. Emission features observed at $\lambda_\textrm{obs}\approx 5533$, $7211$ and $7428\,\si{\angstrom}$ associated with the [OII] doublet, H$\beta$, and [OIII]$_{\lambda5008}$ transitions, respectively, yield a redshift $z=0.4833\pm0.0004$ in agreement with the preliminary estimate of \citet{Cucchiara2016}. 
At this redshift we also observe a low significance feature at the expected location of H$\gamma$. Line properties were derived by fitting the lines with Gaussian functions using the \texttt{specutils} package in \texttt{Python}.

\subsubsection{Radio Observations}

Radio observations were carried out with the Karl J. Jansky
Very Large Array (JVLA) starting at $\sim T_0+1$ d (PI: Berger; project code: 15A-235) with the array in the B configuration. The observations were taken in the X-band, with a central frequency 10 GHz and a bandwidth of 2 GHz. The time on source was 47 minutes. Data were downloaded from the National Radio Astronomical Observatory (NRAO) online archive, and processed locally with the JVLA
CASA pipeline v1.3.2 running in CASA v4.7.2.
We followed the same procedure described in 
\citet{Ricci2020} using galaxies 3C48 and 
J2203+3145 as primary and phase calibrators, respectively. We do not detect a radio transient coincident with the enhanced XRT position with a flux density upper limit $<18$ $\mu$Jy.

\begin{table*}
 	\centering
 	\caption{Log of Optical, nIR, and X-ray Observations of GRB 200522A. Upper limits correspond to a $3\sigma$ confidence level. 
 	}
 	\label{tab: observations_200522A}
 	\begin{tabular}{lccccccc}
    \hline
    \hline
   \multicolumn{8}{c}{\textbf{X-ray Observations}}  \\
   \hline
   \textbf{Start Date (UT)}  & \textbf{$\delta T$ (d)} & \textbf{Telescope} & \textbf{Instrument} &  & \textbf{Exposure (s)} & \textbf{Flux ($10^{-13}$ erg cm$^{-2}$ s$^{-1}$)} &\textbf{Flux Density$^{a}$ ($\mu$Jy)}\\
    \hline
 2020-05-22 11:48:14    & 0.005  & \textit{Swift} & XRT & &   233  & $46\pm10$  & $0.34\pm0.08$ \\
2020-05-22 12:46:40   & 0.045  & \textit{Swift} & XRT &  & 494  & $17\pm5$   & $0.13\pm0.04$\\
      2020-05-22 12:54:57     & 0.051 & \textit{Swift} & XRT &        & 875 & $19\pm4$   & $0.14\pm0.03$\\
   2020-05-22 14:41:21    & 0.12  & \textit{Swift} & XRT & &  2113  & $4.1\pm1.2$    & $0.031\pm0.009$ \\
  2020-05-22 19:41:20     &  0.33 & \textit{Swift} & XRT &  & 4942   &  $1.8\pm0.6$ &  $0.013\pm0.005$\\
   2020-05-23 14:15:51    & 1.1 & \textit{Swift} & XRT &  & 3984  & $1.4\pm0.6$  & $0.010\pm0.004$ \\
     2020-05-25 03:04:32      & 2.6 & \textit{Swift} & XRT &  &  4834  & $<1.1$  & $<8.1\times10^{-3}$  \\
 2020-05-28 02:20:56   & 5.6 & \textit{Chandra} & ACIS-S3 &    &  14890 & $0.13^{+0.05}_{-0.04}$  & $(9.4\pm 2.9)\times 10^{-4}$ \\
2020-06-15 09:15:25 & 23.9 & \textit{Chandra} & ACIS-S3 &    &  57150 &  $<0.03$ &$<2.2\times10^{-4}$ \\
 \hline
 \hline
    \multicolumn{8}{c}{\textbf{Optical/nIR Photometry}}  \\
       \hline
   \textbf{Start Date (UT)}  & \textbf{$\delta T$ (d)} & \textbf{Telescope} & \textbf{Instrument} & \textbf{Filter} & \textbf{Exposure (s)} & \textbf{AB Mag} &\textbf{Flux Density$^{a}$ ($\mu$Jy)}\\ 
    \hline
    2020-05-22 11:49:02 &  0.005 & \textit{Swift} & UVOT & \textit{wh} &  147 & >20.5 & $<25$  \\
    2020-05-22 12:50:03       &  0.048 & \textit{Swift} & UVOT &\textit{u}  & 197  &>21.1  & $<15$ \\
    2020-05-22 12:57:21        &  0.053 & \textit{Swift} & UVOT & \textit{wh} & 197  & >20.8 & $<19$ \\  
   2020-05-22 4:41:43        &  0.19 & \textit{Swift} & UVOT &\textit{u}  & 295  &>21.3  & $<13$ \\   
    2020-05-24 14:36:32   & 2.12 & Gemini-N & GMOS  & \textit{r} & 630 & >22.3  & $<4.6$ \\
    2020-05-25 14:29:19   & 3.12 & Gemini-N & GMOS  & \textit{r} & 720 &$26.0\pm 0.4$ & $0.15^{+0.07}_{-0.05}$  \\
   2020-05-25 23:03:26    & 3.47 & \textit{HST}  & WFC3  &  \textit{F125W}   &  5224  & $24.90\pm0.08$ & $0.41\pm0.02$ \\
    2020-05-26 02:14:16   & 3.61 & \textit{HST}  & WFC3  & \textit{F160W}   &  5224 &$24.71\pm0.07$  & $0.48\pm0.03$ \\
    2020-05-31 14:25:16   & 9.12 & Gemini-N & GMOS  & \textit{r} & 720 & reference & -- \\
    2020-06-07 19:18:57 & 16.32 & \textit{HST}  & WFC3  & \textit{F125W}   & 4824  & >27.2 & $<0.05$ \\
   2020-07-16 15:55:22  & 55.17 & \textit{HST}  & WFC3  & \textit{F125W}   & 5224  & reference & -- \\
   2020-07-16 19:06:10   & 55.31 & \textit{HST}  & WFC3  & \textit{F160W}   & 5224 & reference  & -- \\
    \hline
 \hline
    \multicolumn{8}{c}{\textbf{Optical Spectroscopy}}  \\
       \hline
   \textbf{Start Date (UT)}  & \textbf{$\delta T$ (d)} & \textbf{Instrument} & \textbf{Grating} & \textbf{$\lambda_\textrm{cen}$}  &   \textbf{Exposure (s)}  & \textbf{Slit Width} & \\
   \hline
2020-06-27 13:22:10  & 36.06 & GMOS &  R400  &  750 nm &  3600 & $1\arcsec$ & \\
        \hline
    \end{tabular}
\begin{flushleft}
    \quad \footnotesize{$^a$ Optical/nIR fluxes were corrected for Galactic extinction due to interstellar reddening $E(B-V)=0.02$ mag \citep{Schlafly2011} using the extinction law by \citet{Fitzpatrick1999,Indebetouw2005}. X-ray fluxes were corrected for Galactic absorption $N_H=2.9\times 10^{20} $ cm$^{−2}$ \citep{Willingale2013}, and converted into flux densities at 1 keV using photon index $\Gamma=1.45$.}
\end{flushleft}
\end{table*}

\subsection{GRB 200522A}

\begin{figure} 
\centering
\includegraphics[width=\columnwidth]{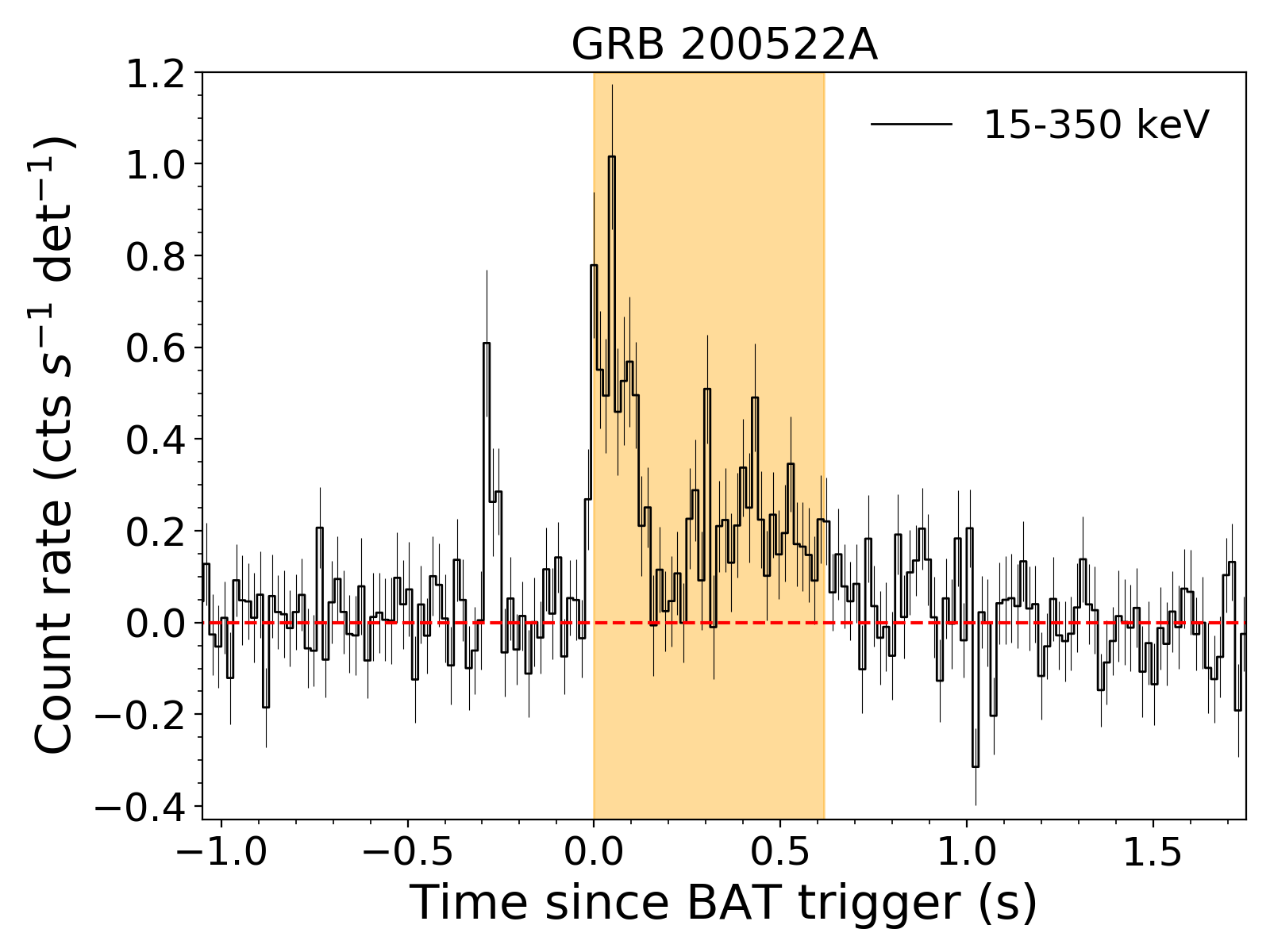}
\caption{BAT mask-weighted lightcurve (15-350 keV) of GRB 200522A with 16 ms binning. Precusor emission is visible $\sim0.25$ s before the BAT trigger. The shaded vertical region marks the $T_{90}$ duration.}
\label{fig: 200522A_bat}
\end{figure}

\subsubsection{Gamma-ray Observations}

\textit{Swift}/BAT was triggered by GRB 200522A on May 22, 2020 at 11:41:34 UT \citep{Evans2020gcn}, hereafter $T_0$ for this GRB. 
The BAT lightcurve, shown in Figure~\ref{fig: 200522A_bat}, 
is multi-peaked with duration $T_{90}=0.62\pm 0.08$ s. A precursor 
\citep{Troja2010} is visible at $T_0-0.25$ s. We find no evidence for EE, and derive an upper limit $<2.2\times10^{-7}$ erg cm$^{-2}$ (15-150 keV) between $T_0+2$ s and $T_0+100$~s.

The BAT GRB Catalogue\footnote{https://swift.gsfc.nasa.gov/results/batgrbcat/} reports that the time-averaged spectrum, from $T_0-0.02$ to  $T_0+0.7$ s, is fit by a power-law with photon index $1.45\pm0.17$ ($\chi^2=39$ for 57 dof). 
For this model, the observed fluence is $f_\gamma = (1.1\pm0.1)\times 10^{-7} $ erg cm$^{-2}$ (15-150 keV),
and we derive an isotropic equivalent gamma-ray energy of $E_{\gamma,\textrm{iso}}=(7.3\pm1.0)\times 10^{49}$ erg (15 keV - 150 keV; rest frame) for a redshift $z\!=\!0.554$ (see \S \ref{sec: 200522A_spectroscopy} and \ref{sec: 200522A_host}). 

\begin{figure*} 
\centering
\includegraphics[width=1.8\columnwidth]{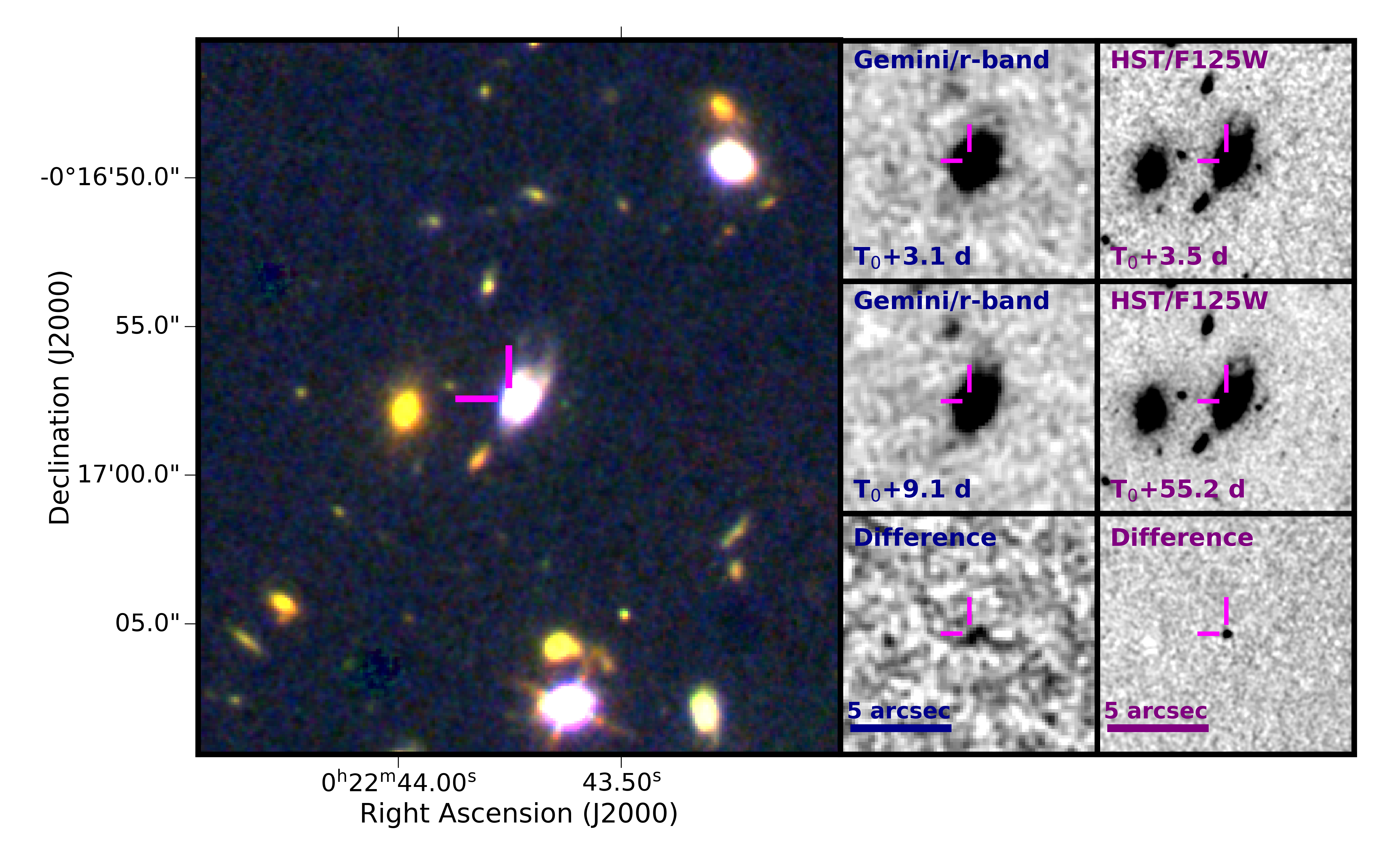}
\caption{\textbf{Left:} RGB image of the field of GRB 200522A: Gemini/\textit{r}-band = blue, \textit{HST}/$F125W$ = green, and \textit{HST}/ $F160W$ = red. The GRB afterglow position is marked by intersecting magenta lines.
\textbf{Center:} Gemini North \textit{r}-band imaging at $T_0+3.1$ d (top) and $T_0+9.1$ d (center). The difference image is displayed in the bottom panel. A weak residual ($\approx\! 3\sigma$) is identified near the bright galaxy's center. The images are smoothed for display purposes. 
\textbf{Right:} \textit{HST} images of GRB 200522A in the $F125W$ filter taken at $T_0+3.5$ d (top) and $T_0+55.2$ d (center). The bottom panel shows the difference image between epochs. A bright, nIR transient is clearly visible.
}
\label{fig: 200522A_RGB}
\end{figure*}

\subsubsection{X-ray Observations}
\label{sec: 200522A_x-rays}

\textit{Swift}/XRT observations were delayed due to the South Atlantic Anomaly, and began at $T_0+406$~s \citep{Evans2020gcn}. The X-ray counterpart was detected at RA, DEC (J2000) = $00^{h}22^m 43^{s}.7$, $-\ang{00;16;59.4}$ with an accuracy of \ang{;;2.2} (90\% CL)\footnote{https://www.swift.ac.uk/xrt\_positions/}. XRT follow-up observations lasted 3 days for a total exposure of 17.5 ks in PC mode. We performed two ToO observations (PI: Troja; ObsIds 22456, 22457, and 23282) with \textit{Chandra}/ACIS-S3 in order to track the late-time evolution of the X-ray lightcurve. During the first epoch ($T_0+5.6$ d), we detect the X-ray afterglow at RA, DEC (J2000) = $00^{h}22^m 43^{s}.74$, $-\ang{00;16;57.53}$ with accuracy $0.5\arcsec$, consistent with the XRT enhanced position. 
A second bright X-ray source lies $\sim 10.4\arcsec$ from the GRB position, and is coincident with the known quasar SDSS J002243.61-001707.8 at redshift $z=1.44862\pm0.00079$ \citep{Krawczyk2013}.  
Due to their proximity, the two sources are not resolved in the 
XRT images and both contribute to the observed X-ray flux.

Analysis of the \textit{Swift} and \textit{Chandra} data was performed using the methods described in \S \ref{sec: 160624A_X-ray_obs}. 
We use our {\it Chandra} observations to characterize the nearby quasar, and 
estimate its contribution to the measured \textit{Swift}/XRT flux. 
Using \texttt{XSPEC}, we derive a photon index $\Gamma=1.53\pm0.14$ (C-stat 162 for 156 dof). This yields a flux $F_X=(6.6\pm0.6)\times 10^{-14}$ erg cm$^{-2}$ s$^{-1}$ (0.3-10 keV).
To constrain the impact of this second source on the XRT observations, we folded the quasar spectrum with the XRT response function to obtain the expected count rate with \textit{Swift}/XRT, $(1.5\pm0.2)\times 10^{-3}$ cts s$^{-1}$ (0.3-10 keV). We subtract this mean count rate from all XRT observations, although the quasar contribution is only significant at late ($>$\,0.3~d)
times.

 The time-averaged XRT/PC mode spectrum from $T_0+400$~s to $T_0+17$~ks 
 is best fit (C-stat=65 for 73 dof) by an absorbed power-law with  photon index $\Gamma=1.45\pm0.18$. We fix the Galactic hydrogen column density to $N_H=2.9\times10^{20}$ cm$^{-2}$ \citep{Willingale2013},
 and include an intrinsic absorption component at the candidate host galaxy's redshift, $z\!=\!0.554$. Our fit sets an upper limit $N_{H,\textrm{int}}\leq 7.4\times10^{21}$ cm$^{-2}$ ($3\sigma$). This yields a time-averaged unabsorbed flux $F_X=(1.2\pm0.2)\times 10^{-12}$ erg cm$^{-2}$ s$^{-1}$, and an ECF of $5.2\times10^{-11}$ erg cm$^{-2}$ cts$^{-1}$.

In our first \textit{Chandra} observation at $T_0+5.6$ d, the afterglow count rate is $(6.8^{+2.6}_{-2.1})\times 10^{-4}$ cts s$^{-1}$ (0.5-7 keV). Using the best fit XRT spectrum, this corresponds to a flux  $(1.3^{+0.5}_{-0.4})\times 10^{-14}$ erg cm$^{-2}$ s$^{-1}$ (0.3-10 keV). 
In the second observation ($T_0+23.9$ d), we detect 2 photons at the GRB position with an estimated background of 0.3 counts yielding a $3\sigma$ upper limit $< 1.6\times 10^{-4}$ cts s$^{-1}$ \citep{Kraft1991}. 
This corresponds to an unabsorbed flux $< 3.0\times10^{-15}$ erg cm$^{-2}$ s$^{-1}$ (0.3-10 keV). The X-ray fluxes from \textit{Swift} and \textit{Chandra} are reported in Table \ref{tab: observations_200522A}.

\subsubsection{Optical/nIR Imaging}
\label{sec: 200522A optical analysis}

The \textit{Swift}/UVOT began settled observations in the \textit{wh} filter at $T_0+448$ s \citep{Kuin2020gcn}. 
Subsequent observations were performed in all optical and UV filters. There was no source detected within the enhanced XRT position. The observations were analyzed using \texttt{HEASoft v6.27.2}. The photometry was performed using circular apertures with a $3\arcsec$ radius and calibrated using the standard UVOT zeropoints \citep{Breeveld2011}; see Table \ref{tab: observations_200522A}.

We imaged the field of  GRB 200522A with GMOS on the 8-m Gemini North telescope (PI: Troja). 
A first set of $r$-band exposures was obtained at $T_0+2.1$~d under poor weather conditions \citep{SimoneGCNimaging}, and repeated at $T_0+3.1$~d. 
A last observation at $T_0+9.1$~d serves as a template for image subtraction. 
The identification of a counterpart is complicated by the presence of a bright galaxy. Image subtraction, using \texttt{HOTPANTS}, between the second ($T_0+3.1$ d) and third ($T_0+9.1$ d) epochs finds a weak ($\approx$3\,$\sigma$) residual source within the 
\textit{Chandra} localization. 
By performing aperture photometry on the difference image we 
estimate a magnitude of $r=26.0\pm0.4$ AB, calibrated against nearby SDSS stars. 

\begin{table}
 	\centering
 	\caption{Observations of the candidate host galaxy of GRB 200522A. Magnitudes are not corrected for Galactic extinction $E(B-V)=0.02$\,mag.}
 	\label{tab: observations_host_200522A}
 	\begin{tabular}{lccc}
    \hline
    \hline
 \textbf{Instrument} & \textbf{Filter} & \textbf{Exp. (s)} & \textbf{AB Mag} \\
    \hline
LDT/LMI & \textit{u} & 1950 & $22.45\pm0.05$\\
LDT/LMI & \textit{g} & 450 & $22.08\pm0.03$\\
Gemini-N/GMOS  & \textit{r} & 720 & $21.30\pm0.04$  \\
LDT/LMI & \textit{i} & 100 &  $21.01\pm0.04$ \\
LDT/LMI & \textit{z} & 600 & $20.93\pm0.03$ \\
Gemini-N/NIRI  & \textit{Y}   &  540 & $20.76\pm  0.10$ 
\\
\textit{HST}/WFC3  & \textit{F125W}   &  5224 &  $20.897\pm0.004$  \\
\textit{HST}/WFC3  & \textit{F160W}   &  5224 & $20.712\pm0.004$  \\
Gemini-N/NIRI  & \textit{$K_s$}   & 180  & $20.88\pm0.17$  \\
     \hline
    \end{tabular}
\end{table}

Near-IR imaging was carried out with the \textit{HST}/WFC3 using the $F125W$ and $F160W$ filters at three epochs (PI: Berger; ObsId 15964): $T_0+3.5$, $T_0+16.3$, and $T_0+55.2$~d. The data was processed using \texttt{sndrizpipe} to a final pixel scale of $0.06\arcsec$/pix. Image subtraction, using \texttt{HOTPANTS}, between the first ($T_0+3.5$ d) and third ($T_0+55.2$ d) epoch uncovers a significant residual source in both filters, at a location consistent with the optical and X-ray positions (see Figure \ref{fig: 200522A_RGB}). The absolute position of the nIR transient is RA, DEC (J2000) = $00^{h}22^m 43^{s}.737$, $-\ang{00;16;57.481}$ with a $1\sigma$ uncertainty of $0.07\arcsec$ (tied to SDSS DR12).
We interpret this source as the optical/nIR counterpart of GRB 200522A, 
as reported in \citet{OConnor2020gcn}. 
Aperture photometry was performed on the residual image 
and calibrated using the tabulated zeropoints. The magnitudes are listed in Table \ref{tab: observations_200522A}.
There are no significant residuals detected at the afterglow location in the $F125W$ difference image between the second ($T_0+16.3$ d) and third epochs. Following the procedure outlined in \S \ref{sec: 160624A_optical_analysis}, we inject artificial point sources to determine an upper limit $F125W>27.2$ AB mag at the afterglow position.

An independent analysis of the \textit{HST} data was recently reported by \citet{Fong2020}, confirming our detection of the optical/nIR counterpart. 
The analysis of \citet{Fong2020} reports a source brighter by $\approx0.4$ mag in the $F125W$ filter, which we can reproduce by using a different template image derived by combining the two epochs at $T_0+16$ and 55 days using the \texttt{astrodrizzle} package. In our work we adopt instead the results derived using the single epoch at $T_0+55$ d, available for both the $F160W$ and $F125W$ filters, as small variations in the nearby galaxy's nucleus may affect the photometry and the measured color. 
We caution that our error bars do not include a systematic uncertainty accounting for possible variability of the galaxy's nucleus. However, we verified that all our conclusions also hold for the
alternative result of a slightly brighter transient (see \S \ref{sec: 200522A_afterglow_overview}).

Late-time optical and nIR images were acquired to characterize the host galaxy's properties. Observations were carried out in the \textit{ugiz} filters with the LDT/LMI on July 30, 2020, and in the \textit{YK$_s$} filters with 
Gemini/NIRI on July 17, 2020. Data were reduced following the same procedures described in \S \ref{sec: 160624A_optical_analysis}, and photometry was calibrated using nearby sources from SDSS and the United Kingdom Infrared Telescope Infrared Deep Sky Survey \citep[UKIDSS;][]{Lawrence2007}. The results are listed in Table \ref{tab: observations_host_200522A}.

\subsubsection{Optical Spectroscopy} 
\label{sec: 200522A_spectroscopy}
\begin{figure} 
\centering
\includegraphics[width=\columnwidth]{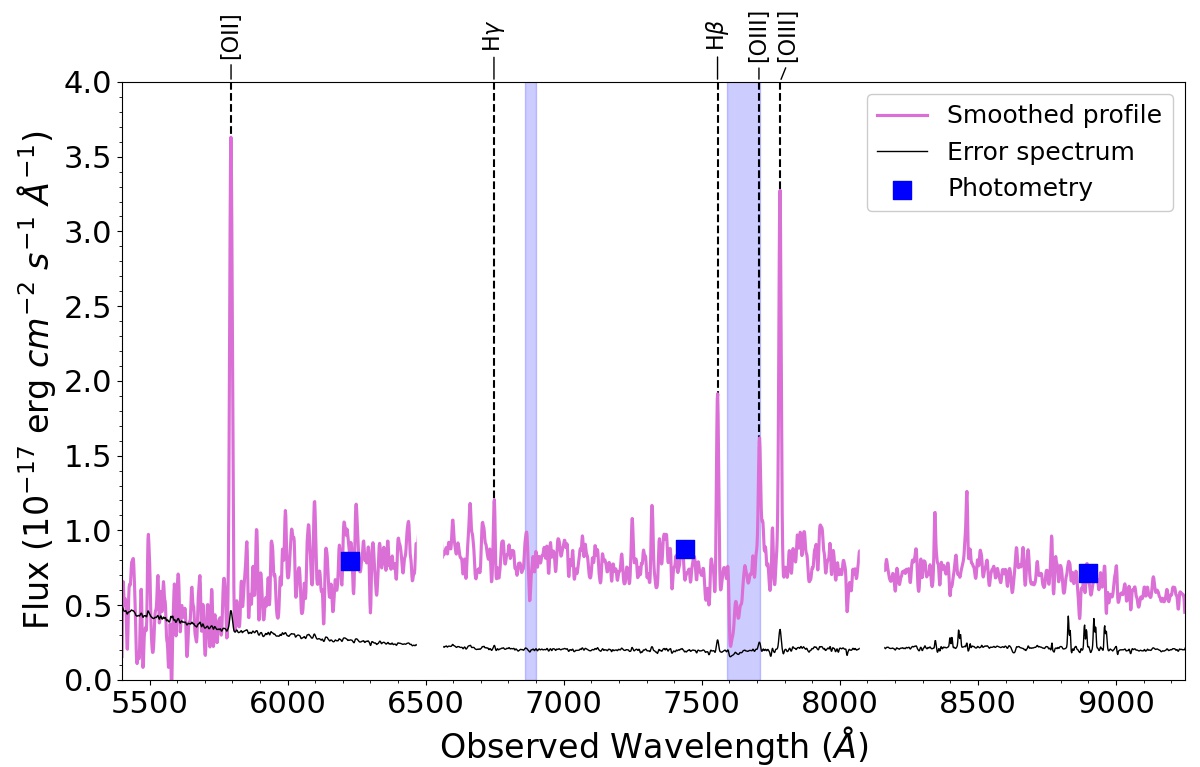}
\caption{
Gemini North GMOS spectrum of the host galaxy of GRB 200522A at $z=0.5541\pm 0.0003 $. The spectrum has been smoothed with a Gaussian kernel, presented in purple, and the error spectrum is shown in black. The host galaxy's photometry is shown as blue squares, corresponding to the Gemini/\textit{r}-band and LDT/\textit{i}- and \textit{z}-bands. Telluric absorption regions are marked by blue bands. The location of detected emission features are indicated by dashed black lines. There are chip gaps at $6500\si{\angstrom}$ and $8100\si{\angstrom}$.
}
\label{fig: 200522A_host_spectrum}
\end{figure}

We obtained a spectrum of the putative host galaxy using GMOS on the 8-m Gemini North telescope (PI: Troja) on June 27, 2020 ($T_0 + 36$ d). 
We performed a set of $6\times600$ s exposures using the R400 grating with central wavelength $\approx$740 nm.
The resulting combined spectrum is shown in Figure \ref{fig: 200522A_host_spectrum}. We identify emission features at $\lambda_\textrm{obs}\approx 5795, 6747, 7556, 7782, 7708 \,\si{\angstrom}$ which are associated with the [OII] doublet,  H$\gamma$, H$\beta$, [OIII]$_{\lambda4959}$, and [OIII]$_{\lambda5008}$ transitions, respectively. This yields a redshift $z=0.5541 \pm 0.0003$, in agreement with our preliminary estimate \citep{SimoneGCN}. Line properties were derived through the methods outlined in \S \ref{sec: 160624A spectrum}.

\section{Methods}
\label{sec: methods}
\subsection{Afterglow Modeling}
\label{sec: model_fitting_methods}


We model the observed afterglows within the standard fireball model \citep{Meszaros1997,Sari1998,Wijers1999,Granot2002}, described by a set of five parameters:
the isotropic-equivalent kinetic energy $E_\textrm{0}$,
the circumburst density $n_0$, the fraction of energy in magnetic fields $\varepsilon_B$ and in electrons $\varepsilon_e$, and the slope $p$ of the electron energy distribution $N(E) \propto E^{-p}$. 
We assume that the environment surrounding the binary merger has a uniform density profile, consistent with the interstellar medium (ISM). 
Three more parameters account for the outflow's collimated geometry: 
the jet core width $\theta_c$, the observer's viewing angle $\theta_v$, 
and the jet's angular profile. 
We apply two angular profiles: (i) a uniform (tophat) jet profile with $\theta_v$\,$\approx$\,0 and (ii) a Gaussian function in angle from the core described by $E(\theta)=E_0 \exp(-\theta^2/2\theta_c^2)$ for $\theta\leq\theta_w$, where $\theta_w$ is the truncation angle of the Gaussian wings. 
The beaming corrected kinetic energy, $E_j$ is given by $E_j=E_0(1-\cos\theta_c)\approx E_0\theta_c^2/2$ for a top-hat jet and $E_j\approx E_0\theta_c^2[1-\exp(-\theta^2_w/2\theta_c^2)]$ for a Gaussian angular profile. 
We include the effect of intrinsic dust extinction assuming a \citet{Fitzpatrick1999} reddening law parametrized by $R_V=A_V/E(B-V)=3.1$.

We utilize a Bayesian fitting method in conjunction with the {\tt afterglowpy} software\footnote{\url{https://github.com/geoffryan/afterglowpy}}, described in \cite{Ryan19}, to determine the GRB jet parameters. We apply the \textsc{emcee} (version 2.2.1; \citealt{emcee}) Python package for Markov-Chain Monte Carlo (MCMC)  analysis. 
The independent priors for each parameter were uniform for 
$E(B-V)$ =[$0, 1$], $\theta_c$ =[$0, \pi/4$] and $p$=[$2, 5$], and  log-uniform for $\log E_0$=[$48, 55$], $\log n_0$=[$-6,2$], $\log \varepsilon_B$=[$-6,-0.5$], and $\log\varepsilon_e$=[$-6,-0.5$]. 
We have restricted $\varepsilon_e,\varepsilon_B\!<\!1/3$, as without this requirement both $\varepsilon_e$ and $\varepsilon_B$ approach unphysical values of unity.
For the Gaussian jet profile, we adopt the uniform priors $\theta_v$ =[$0, \pi/4$] and $\theta_w$ =[$0.01, \min(\pi/4,12\theta_c$)]. 
 Each fit used an ensemble MCMC sampler that employed 300 walkers for 100,000 steps with an initial burn-in phase of 25,000 steps, yielding $2.25\times10^7$ posterior samples. Additional details of the methods can be found in \citet{Troja18, Piro19, Ryan19}.

We compare models by evaluating their predictive power with the Widely Applicable Information Criterion \citep[WAIC, also known as the Watanabe-Akaike Information Criterion;][]{Watanabe:2010}. The WAIC score is an estimate of a model's expected log predictive density (\emph{elpd}): roughly the probability that a new set of observations would be well described by the model's fit to the original data. A model with a high \emph{elpd} (and WAIC score) produces accurate, constraining predictions. A model with a low \emph{elpd} (and WAIC score) produces predictions which are inaccurate or not constraining. This naturally penalizes over-fitting: over-fit models typically show wide variability away from the observations which leads to poor predictive power and a low WAIC score.  The WAIC score estimates the \emph{elpd} by averaging the likelihood of each observation over the entire posterior probability distribution.  As a model comparison tool, this incorporates the uncertainties in the model parameters (unlike the reduced $\chi^2$ or the Akaike Information Criterion) and does not require the posterior probability distribution to be normal (unlike the the Deviance Information Criterion).

We compute the WAIC score following \citet{Gelman:2013} using the recommended $p_{\mathrm{WAIC\ \! 2}}$ estimate for the effective number of parameters. The WAIC score is computed at every datapoint and the total WAIC score is the sum of these contributions. In order to compare the WAIC score of two different models, we compute the standard error, $\sigma_{SE}$, of their difference, $\Delta$WAIC$_{\mathrm{elpd}}$ \citep{Vehtari:2015}. One model is favored over another if it has a higher overall WAIC score and the difference between the WAIC scores $\Delta$WAIC$_{\mathrm{elpd}}$ is significantly larger than its uncertainty $\sigma_{SE}$. As $\sigma_{SE}$ can underestimate the true standard deviation of $\Delta$WAIC$_{\mathrm{elpd}}$ by up to a factor of two \citep{Bengio:2004}, we report a conservative significance on the WAIC score difference using $\sigma_{\Delta\textrm{WAIC}}\approx2\sigma_{SE}$.

\subsection{Kilonova Modeling}
\label{sec: kilonova_models}

\subsubsection{Empirical Constraints}
\label{sec: simple BB}

Optical and infrared observations constrain the properties of possible kilonovae associated with each GRB.
In the case of GRB~160624A, no optical or nIR counterpart was detected and kilonova models are directly constrained by our photometric upper limits. 
GRB~200522A presents instead a complex phenomenology, characterized by a bright X-ray afterglow and an optical/nIR counterpart with a red color
($r-H \approx$1.3, Table \ref{tab: observations_200522A}).
Such red color could be the result of dust along the sightline or the telltale signature of a kilonova. 
In order to constrain the contribution of the latter, we add to our afterglow fit an additional thermal component.  

\citet{Korobkin2020} demonstrated that a simple analytical fit to kilonova lightcurves can lead to order of magnitude uncertainties in the inferred ejected mass, depending on the unknown geometry of the system.
We therefore use a different approach which combines empirical constraints and detailed numerical simulations. 
At the time of our optical/nIR observations ($\sim 3.5$ d), the kilonova component is roughly described by a simple blackbody.
Therefore, we use a parameterized blackbody component, included in addition to the standard forward shock (FS) emission, to determine the possible thermal contribution from a kilonova. 
The range of fluxes allowed by this fit is then compared to an extensive suite of simulated kilonova lightcurves (\S \ref{sec: KN_grid}) in order to derive the ejecta properties. 
The blackbody component is described by two parameters, its temperature $T$  and emission radius $R$, with uniform priors between [0-8,000 K] and [0-5$\times 10^{15}$ cm], respectively. The upper limit on radius is chosen to prevent a superluminal expansion velocity. Since our optical and nIR data are nearly coeval, we assume that there is negligible temporal and spectral evolution between observations.

\subsubsection{Constraints on Kilonova Ejection Properties}
\label{sec: KN_grid}

We compare optical and infrared constraints to simulated kilonova lightcurves of varying input parameters, consistent with a wide range of plausible ejecta morphologies, compositions, masses, and velocities. For this study, we use a grid of two-component kilonova models from the LANL group (Wollaeger et al., in prep). This data set was previously used in \citet{Thakur2020} to constrain ejecta parameters for GW190814. Simulations include time-dependent spectra, as early as three hours post-merger, which are subsequently converted to lightcurves for various filters. We simulate kilonovae with \texttt{SuperNu} \citep{Wollaeger2014}, a multi-dimensional, multi-group Monte Carlo transport code, which has previously been used in a wide range of kilonova studies \citep{Wollaeger2018, Wollaeger2019, Even2020, Korobkin2020, Thakur2020}. Our simulations rely on the WinNet nucleosynthesis network \citep{Winteler2012} to simulate heating from radioactive decay in addition to the latest LANL opacity database \citep{Fontes2020}. We consider a full set of lanthanide opacities, while uranium acts as a proxy for all actinide opacities. 

We model kilonovae with two ejecta components: dynamical ejecta including heavy r-process elements and wind ejecta emanating from the resultant accretion disk surrounding the remnant compact object. We consider two disparate wind models, representing ejecta with either high-latitude ($Y_e = 0.37$) or mid-latitude ($Y_e = 0.27$) compositions, both having negligible lanthanide contributions.  Wind ejecta assumes either a spherical or ``peanut-shaped'' morphology, while dynamical ejecta remains toroidal. These models correspond to the TS and TP morphologic profiles in \citet{Korobkin2020}. 

The grid of models includes a range of mass and velocity parameters, in addition to two morphologies and two wind compositions. Both the dynamical and wind ejecta span five possible masses: 0.001, 0.003, 0.01, 0.03, 0.1 $M_\odot$. We ascribe one of three possible velocity distributions to both the dynamical and wind ejecta components, with median velocities of either 0.05$c$, 0.15$c$, or 0.3$c$, corresponding to maximum ejecta velocities of 0.1$c$, 0.3$c$, and 0.6$c$. The grid spans the anticipated range of ejecta properties expected from numerical simulations and observations of GW170817 \citep{korobkin12, Kasen2017, cote18, Metzger2019, Kruger2020}. Each multi-dimensional simulation computes kilonova emission for 54 different polar viewing angles. Our axisymmetric simulations report spectra and lightcurves for separate viewing angles, distributed uniformly in sine of the polar angle from edge-on, on-axis viewing angles (0\degree) to edge-off (180\degree). Including all viewing angles and kilonova properties, we have 48,600 different sets of time-dependent spectra to compare to optical and infrared observations.

We limit our simulation grid to kilonovae observed on-axis, when considering optical and infrared observations in conjunction with a GRB, observed with viewing angle $\theta_v$\,$\approx$\,0. Our on-axis simulations consider viewing angles from $0\degree$ to $15.64\degree$. We then compare the 900 on-axis kilonova simulations to optical and near-infrared observations. We restrict plausible kilonova parameters to the range of simulated lightcurves consistent with observations, either lightcurves dimmer than the measured upper limits (see \S\ref{sec: 160624A_kilonova}) or lightcurves residing in the range of inferred kilonova emission from analytic fits (see \S\ref{sec: 200522A_kilonova}).

\subsection{Galaxy SED Modeling}
\label{sec: SED fitting}

We compare the observed photometry to a range of synthetic spectral energy distributions (SEDs) generated with the flexible stellar population synthesis (FSPS) code \citep{Conroy2009}. We adopt the same models used in \citet{Mendel2014} to describe SDSS galaxies: a \citet{Chabrier2003} initial mass function (IMF) with integration limits of 0.08 and 120 $M_{\odot}$ (\texttt{imf\_type = 1}); an intrinsic dust attenuation using the extinction law of \citet[][\texttt{dust\_type = 2}]{{Calzetti2000}}; a smoothly declining star-formation history characterized by an e-folding timescale, $\tau$. We apply a delayed-$\tau$ model (\texttt{sfh=4}) for the star-formation history. 
The contribution of nebular emission is computed using the photoionization code \textsc{Cloudy} \citep{Ferland2013} and added to the spectrum.
These choices result in 5 free parameters: the total stellar mass formed $M$, the age $t_{\rm age}$ of the galaxy, the e-folding timescale $\tau$, the intrinsic reddening $E(B-V)$, and the metallicity $Z_*$. From these parameters, we derive the stellar mass, $M_*$ = $\xi\,M$, where $\xi$ is the ratio of the surviving stellar mass to the formed mass, and the star-formation rate, SFR, computed as: 
\begin{equation}
\textrm{SFR} (t_{\rm age}) = 
M~\frac{1}{\tau^2\,\gamma(2,t_\textrm{age}/\tau)}~
t_{\rm age}\,e^{-\frac{t_{\rm age}}{\tau}},
\end{equation}
where $\gamma(a,x)$ is the lower incomplete gamma function. 
The mass-weighted stellar age, $t_m$, is then derived as:
\begin{align}
    t_m = t_\textrm{age} - \frac{\int^{t_\textrm{age}}_0 t \times \,\textrm{SFR}(t) \, dt}{\int^{t_\textrm{age}}_0 \textrm{SFR}(t)\, dt}.
\end{align}

We sampled the posterior probability density function of these parameters by using the affine-invariant ensemble MCMC sampler implemented in the \textsc{emcee} package \citep{emcee}. We adopted uniform priors in log $t_{\rm age}$, log $\tau$, log $Z$, $E(B-V)$ over the same parameter range as \citet[][cf. their Table 2]{Mendel2014}, and ran each fit with 128 walkers for 4096 steps, dropping the initial 100,000 steps as an initial burn-in phase and generating $\approx$400,000 posterior samples. The MCMC walkers were initialized near the maximum of the posterior, calculated through optimization of the likelihood function. Fits were performed with the \textsc{prospector} code \citep{Johnson2019}, customized to use our chosen cosmology.

\section{Results}
\label{sec: results}

GRB 160624A and GRB 200522A are two short duration GRBs located at a similar distance ($z\!=\!0.483$ and $0.554$, respectively), which 
display very different properties in their observed emission. 
GRB 160624A is characterized by a bright, short-lived X-ray afterglow that is no longer detected after a few hundreds seconds. The faint afterglow and lack of any optical/nIR counterpart allow for stringent constraints on kilonova emission from the deep Gemini and \textit{HST} observations (see \S \ref{sec: 160624A section}). 
GRB 200522A displays instead a bright and long-lived counterpart.
The red color of its optical/nIR emission (\textit{r-H}\,$\approx1.3$ mag) represents tantalizing evidence for a kilonova, but its interpretation is complicated by the uncertain contribution of the standard afterglow. The burst location, close to the galaxy's center, and evidence for active star-formation suggest that extinction along the sightline could also contribute to the observed color (see \S \ref{sec: 200522A section}). 

\begin{figure}
\centering
\includegraphics[width=\columnwidth]{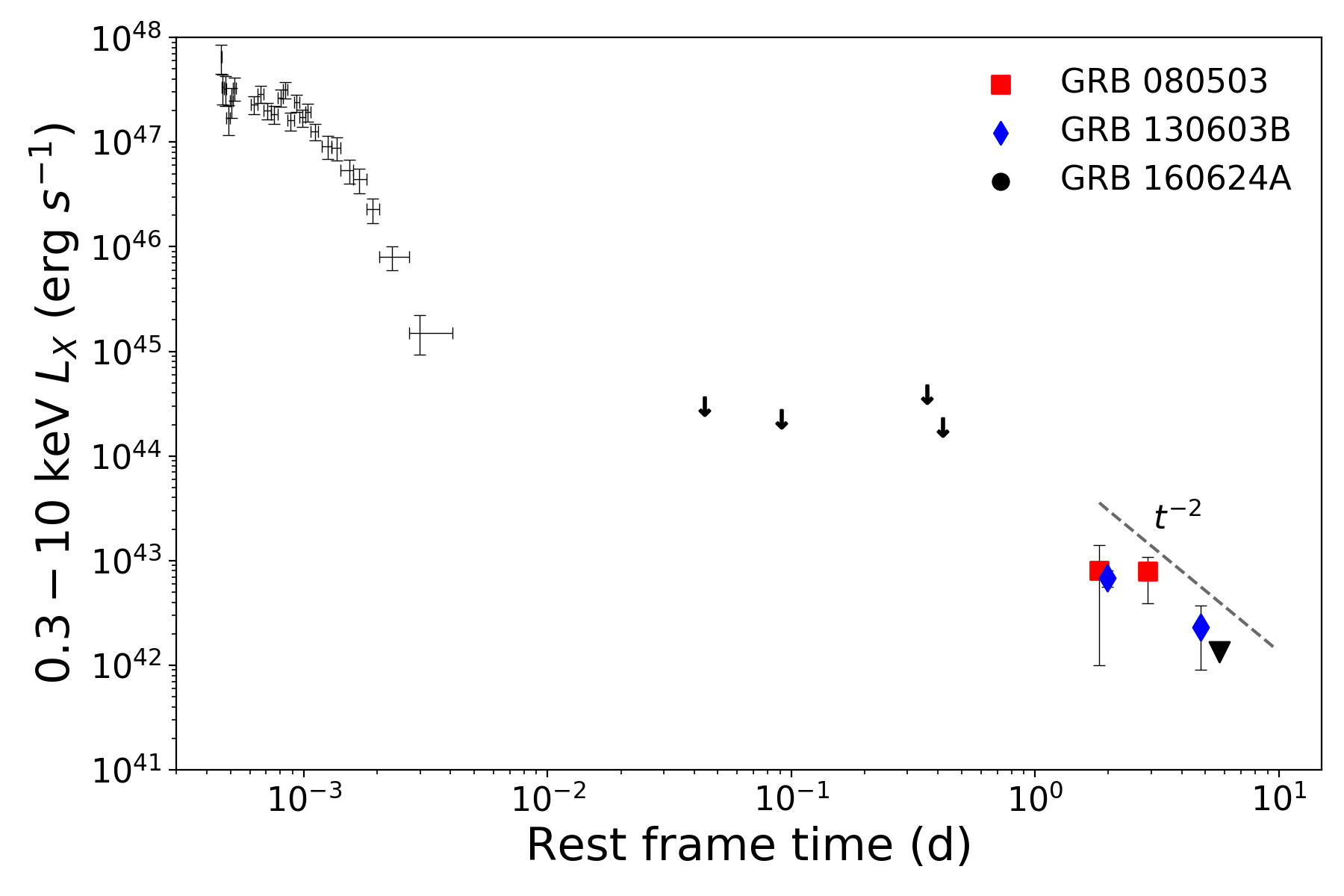}
\caption{Rest frame X-ray lightcurve of GRB 160624A in the 0.3-10 keV band. For reference, we show the observed late-time X-ray excess in GRBs 080503 \citep{Perley2009} and 130603B \citep{Fong2014}. The dashed line corresponds to the decay ($\propto t^{-2}$) of predicted late-time X-ray excess from a long-lived magentar central engine. Upper limits ($3\sigma$) are represented by downward arrows for XRT and downward triangles for \textit{Chandra}.
}
\label{fig: 160624A_lightcurve}
\end{figure}

\subsection{GRB 160624A}
\label{sec: 160624A section}
\subsubsection{Afterglow Properties}
\label{sec: 160624A_afterglow}

\begin{figure} 
\centering
\includegraphics[width=\columnwidth]{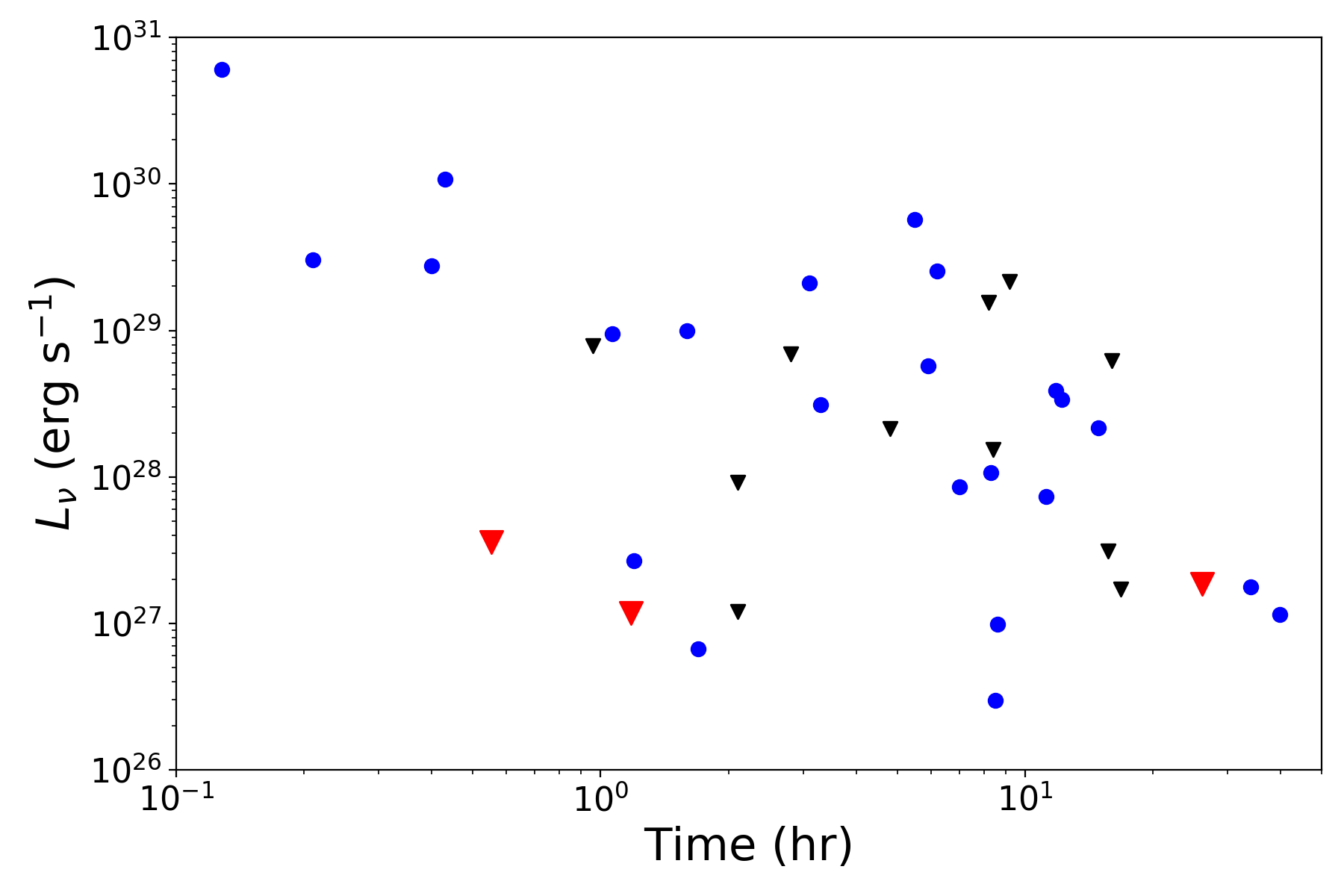}
\caption{
Optical upper limits (black triangles) or detections (blue circles) in the \textit{r}-band for sGRBs at their time of first observation, measured since the GRB trigger.  The three early Gemini/GMOS \textit{r}-band upper limits for GRB 160624A are shown (red triangles) in comparison to the rest of the population of sGRBs with measured redshift. The measurements are corrected for Galactic extinction \citep{Schlafly2011}.
}
\label{fig: SGRB opt limits}
\end{figure}

As shown in Figure \ref{fig: 160624A_lightcurve}, GRB~160624A displays a bright and rapidly fading X-ray afterglow. This feature is common
among sGRBs and is often interpreted as long-lived activity of the central engine \citep[e.g.,][]{Rowlinson2013}. No evidence for a standard FS component is found by deep optical and X-ray follow-up observations. 
At early times ($<$2 hrs), this event is characterized by the deepest available optical limits for a sGRB \citep[see also, e.g.,][]{Sakamoto2013,Fong2015}. These observations would have detected nearly all the known sGRB optical afterglows, with the only exception of GRB~090515, and place some of the tighest constraints
on the optical luminosity of sGRBs (Figure \ref{fig: SGRB opt limits}). 

Using \texttt{afterglowpy}, we explore the range of afterglow parameters allowed by the broadband upper limits. 
We fixed $p=2.2$, and left the other parameters ($E_0$, $n_0$, $\varepsilon_B$, $\varepsilon_e$) free to vary. 
Although low density ($\approx$ $10^{-3}$ cm$^{-3}$) solutions are favored, $\sim15\%$ of the allowed models are consistent with $n_0\gtrsim 1$ cm$^{-3}$. The faintness of the afterglow therefore does not necessarily imply a rarefied environment. 


Our \textit{Chandra} observation sets an upper limit to the X-ray luminosity  $L_X\lesssim1.1\times10^{42}$ erg s$^{-1}$ (0.3-10 keV) at $T_0+5.9$~d (rest frame). This limit is compared in Figure \ref{fig: 160624A_lightcurve} to the late-time X-ray excess detected for short GRBs 080503 \citep{Perley2009} and 130603B \citep{Fong2014}, 
which is often attributed to a long-lived and highly magnetized NS remnant  \citep[e.g.,][]{Gao2013,Metzger2014}. The interaction of the magnetar spin-down radiation with the merger ejecta could power a bright X-ray transient on timescales of a few days after the merger. The predicted peak luminosity is $10^{43}-10^{45}$ erg s$^{-1}$ with a decay following the temporal behavior of the spin-down emission, $L_\textrm{sd}\propto t^{-2}$. In order to be consistent with these models, our non-detection of X-rays favors a newborn NS with a large magnetic field $B\gtrsim10^{15}$ G for ejecta masses $M_\textrm{ej}\gtrsim10^{-2} M_\odot$ \citep{Metzger2014}. Alternatively, the early steep decay of the X-ray afterglow may mark the NS collapse to a BH.

\subsubsection{Kilonova Constraints}
\label{sec: 160624A_kilonova}



\begin{figure*} 
\centering
\includegraphics[width=1.7\columnwidth, trim=100 100 100 100, clip]{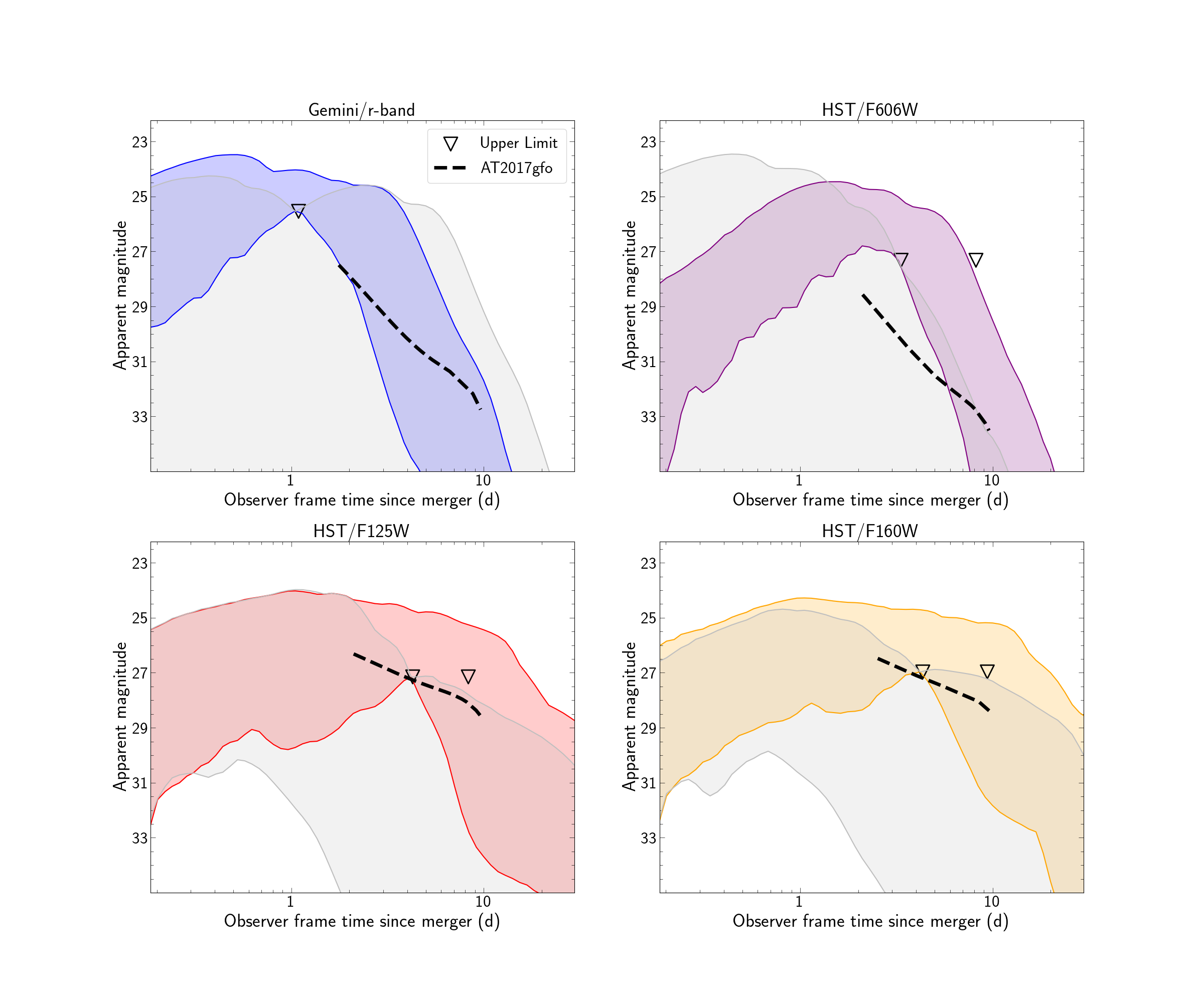}
\caption{ 
Kilonova lightcurves, in the observer frame, from the LANL simulation suite for Gemini/\textit{r}-band (upper left), \textit{HST}/$F606W$ (upper right), \textit{HST}/$F125W$ (lower left), and \textit{HST}/$F160W$ (lower right) filters compared to upper limits (downward triangles) for GRB 160624A. Gray bands mark the magnitude extent of lightcurves consistent with our upper limits, while the colored bands correspond to lightcurves disallowed by observations. Only lightcurves for on-axis viewing angles ($\theta_v\lesssim15.64^\circ$) are considered. AT2017gfo lightcurves (dashed black lines in each panel) are included for comparison.
}
\label{fig: 160624A_kilonova_lightcurve}
\end{figure*}

\begin{figure} 
\centering
\includegraphics[width=1\columnwidth, trim=0 75 0 70, clip]{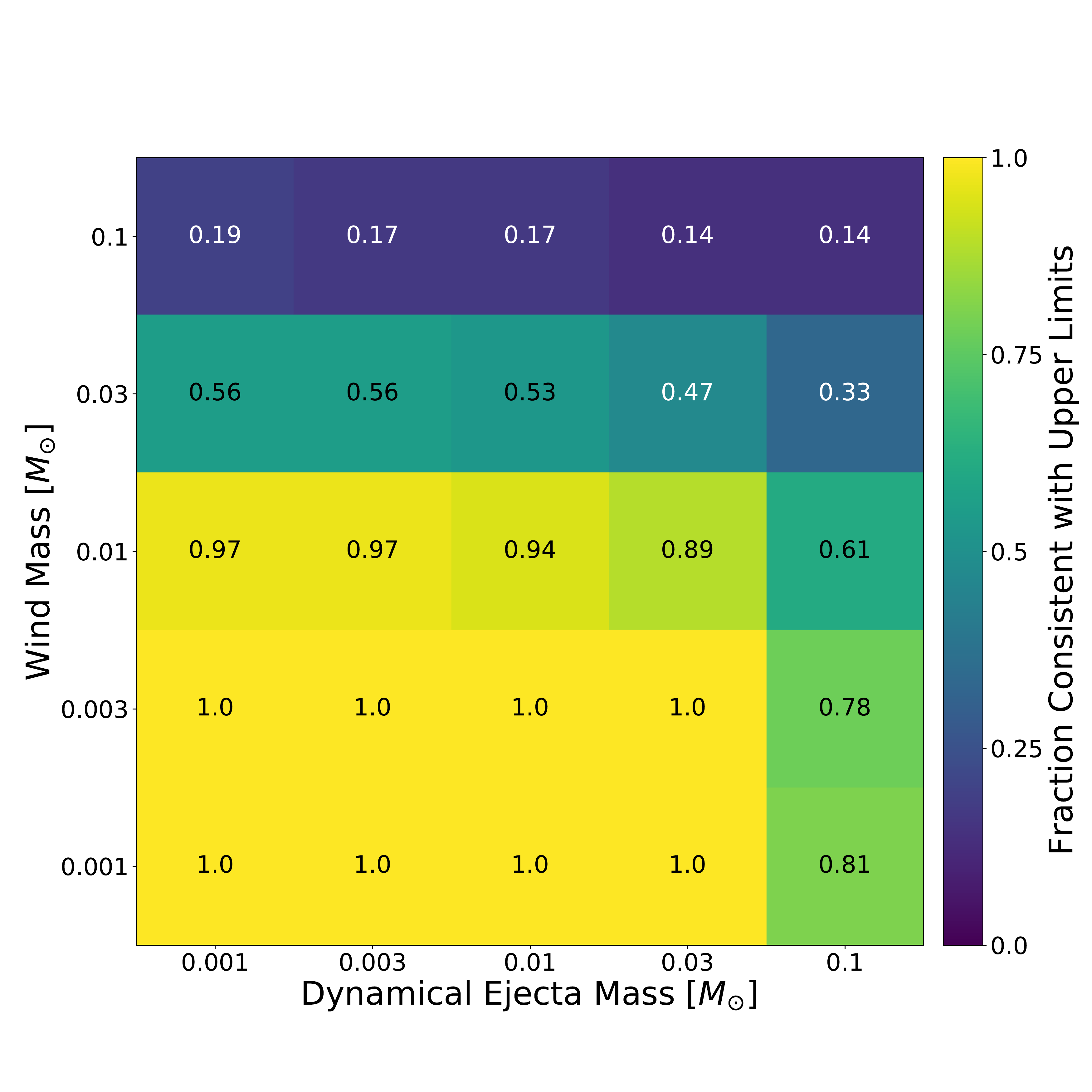}
\caption{Fraction of simulated kilonovae consistent with the observational constraints for GRB 160624A. Our two-component models assume five possible masses for both the dynamical and wind ejecta components. Each square corresponds to the set of 36 lightcurves with the described mass parameters. 
}
\label{fig: 160624A_kilonova_constraints}
\end{figure}

We constrain parameters of a potential kilonova associated with GRB 160624A by comparing optical and infrared upper limits to 900 on-axis kilonova simulations, introduced in \S \ref{sec: KN_grid}. We only consider observations after $T_0+0.125$ d (rest frame), as spectra are not computed prior to this time. Using spectra simulated at various times, we convert kilonova emission to observer frame lightcurves in Gemini/\textit{r}-band, \textit{HST}/$F606W$, \textit{HST}/$F125W$, and \textit{HST}/$F160W$ filters.

The four panels of Figure~\ref{fig: 160624A_kilonova_lightcurve} show the range of simulated lightcurves in each filter. Colored regions indicate the range of lightcurves eliminated by upper limits, while gray regions indicate lightcurves consistent with observations. AT2017gfo lightcurves are included for comparison, and show that \textit{HST}/$F125W$ and \textit{HST}/$F160W$ observations may be sensitive to AT2017gfo-like kilonovae. The AT2017gfo photometry was 
compiled from \citet{Arcavi2017,Cowperthwaite2017,Drout2017,Kasliwal2017,Pian2017,Shappee2017,Smartt2017,Tanvir2017,Troja2017,Utsumi2017,Valenti2017}. 

Our \textit{HST}/$F160W$ observations provide the most stringent constraints on the range of plausible lightcurves, disallowing 30\% of on-axis lightcurves from the simulation grid. Individually, the Gemini/\textit{r}-band upper limit at $\sim 1$ d (observer frame) eliminated 12\% of lightcurves, while the earlier \textit{HST}/$F606W$ and \textit{HST}/$F125W$ observations eliminate 8\% and 23\% of lightcurves, respectively. The \textit{HST}/$F125W$ upper limit at $\sim 8$ d post-merger (observer frame) and the \textit{HST}/$F160W$ upper limit at $\sim 9$ d post-merger (observer frame) only provide redundant constraints, eliminating lightcurves otherwise constrained by the four aforementioned upper limits. The \textit{HST}/$F606W$ upper limit at $\sim 8$ d post-merger (observer frame) places no constraint on the range of kilonova parameters. In total, 31\% of simulated on-axis kilonovae are ruled out by the observational constraints on GRB 160624A. 

Figure~\ref{fig: 160624A_kilonova_constraints} indicates the fraction of simulated lightcurves consistent with observations for all combinations of dynamical and wind ejecta mass. Constraints indicate that high ejecta masses are strongly disfavored, with 86\% of simulations with 0.2 $M_\odot$ total ejecta mass (0.1\,$M_\odot$ of dynamical ejecta + 0.1\,$M_\odot$ wind ejecta) excluded by observational constraints.  High wind ejecta masses ($\geq$ 0.1 $M_\odot$) are disfavored with over 80\% of models disallowed by the upper limits. 
Wind mass dictates lightcurve behavior at lower (optical) wavelengths, while dynamical ejecta mass dominates at higher (nIR) wavelengths. 
However, due to the cosmological distance of this GRB, our reddest filter, \textit{HST}/$F160W$, corresponds only to the rest frame \textit{y}-band.
As a result, our observations can only weakly constrain the dynamical ejecta with 53\% of 0.1 $M_\odot$ models consistent with the data. Nearly all models with ejecta masses below 0.04 $M_\odot$ are consistent with the data. 

For high ejecta masses, we are able to place strong constraints on the range of ejecta velocities. For example, lightcurves with wind ejecta masses of 0.1 $M_\odot$ and low velocities ($\leq 0.15c$) are strongly disfavored. Similarly, models with wind ejecta 0.03 $M_\odot$ and low velocity ($0.05c$) are predominately disfavored, while the majority of high mass models (with velocity $0.3c$) are consistent with the data. Velocity constraints are primarily due to the timing of our observations. For constant mass, higher velocities result in earlier and brighter peak emission \citep[see, e.g.,][]{Thakur2020}. As a result, many high velocity models have dimmed by the time of these observations and cannot be ruled out by the upper limits, while several low wind velocity models coincide with observations and are thus ruled out.

\subsubsection{Environment}
\label{sec: 160624A_host}

The best localization for GRB~160624A is its XRT position with an error radius of 1.7$\arcsec$ (90\% CL). This position intercepts a bright galaxy (Figure \ref{fig: 160624A_RGB}), which we identified as the GRB host galaxy.
Using the XRT localization, the maximum projected physical offset from this host galaxy is $\lesssim 21.5$ kpc (90\% CL). 
Using the \texttt{GALFIT} package \citep{Peng2002}, we fit this galaxy with a
Sersic profile of index $n$=1 and derive an optical half-light radius $R_e \approx 1.1\arcsec$ ($6.8$ kpc). 
Therefore, the maximum host-normalized offset is $R/R_e\lesssim 3.2$.

Following \citet{Bloom2002}, 
we calculate the chance probability $P_{cc}$ using the R-band number counts from \citet{Beckwith2006} for optical observations, and the H-band number counts from \citet{Metcalfe2006} for nIR observations. 
We derive $P_{cc}=$0.03 using the observed magnitude $r=22.18\pm0.02$, and $P_{cc}=$0.02 using \textit{HST}/$F160W=20.566\pm0.004$ (see Table \ref{tab: observations_host}).
We searched the field for other candidate host galaxies. We identify three bright SDSS galaxies at offsets of \ang{;;21}, \ang{;;22.4}, and \ang{;;39} with $P_{cc}\!\approx\! 0.2$, $0.5$, and $0.9$, respectively. There are a few dim extended objects, uncovered in the \textit{HST} observations, at moderate offsets $\gtrsim\! 4 \arcsec$ with $P_{cc}\!\gtrsim\!0.5$. Additionally, a faint 
source with $F160W=26.2\pm0.3$ mag is observed within the XRT position. Due to its faint nature, we cannot determine whether this is a star or a galaxy. 
In the latter case, the source's probability of chance coincidence is $P_{cc}\!\approx\!0.7$. 
Therefore, the association of GRB 160624A with the bright galaxy SDSS J220046.14+293839.3 remains the most likely.

\begin{figure} 
\centering
\includegraphics[width=\columnwidth]{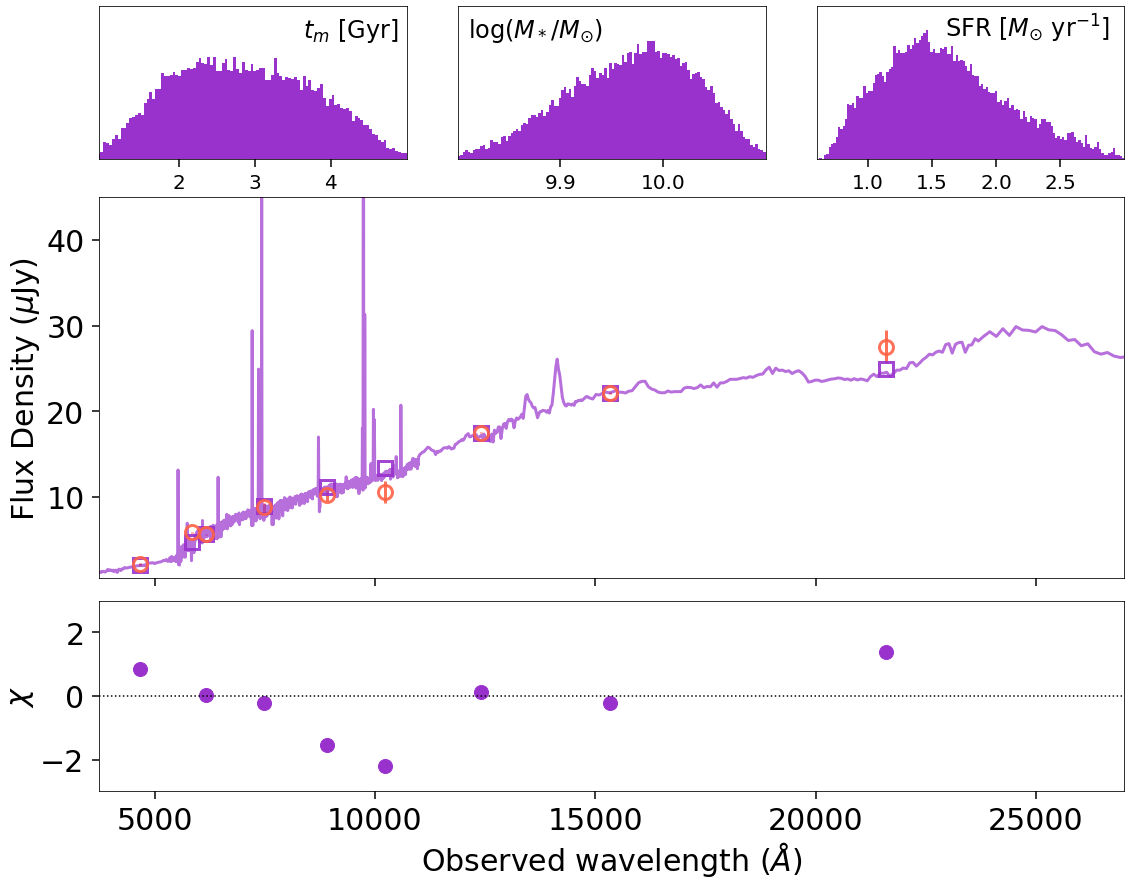}
\vspace{-0.5cm}
\caption{The best fit model spectrum (solid line) and photometry (squares) describing the host galaxy SED for GRB 160624A compared to the observed photometry (circles), which is corrected for Galactic extinction. The bottom panel shows the fit residuals. The stamps (top) demonstrate posterior distributions for $t_m$, $\log(M_*/M_\odot)$, and SFR.
}
\label{fig: 160624A_SED}
\end{figure}

We characterize the putative host galaxy's properties by modeling the optical and nIR SED (Table \ref{tab: observations_host}) as outlined in \S \ref{sec: SED fitting}. The result is shown in Figure \ref{fig: 160624A_SED}. The best fit parameters describing the galaxy are: an intrinsic extinction $E(B-V)=0.13\pm0.06$ mag, a near solar metallicity $Z_*/Z_\odot=0.9\pm0.3$, an $e$-folding time $\tau=1.4^{+0.9}_{-0.6}$ Gyr, a stellar mass $\log(M_*/M_\odot)=9.97^{+0.06}_{-0.07}$, an old stellar population $t_m=2.8^{+1.0}_{-0.9}$ Gyr, and a moderate star-formation rate SFR $=1.6^{+0.6}_{-0.4}$ $M_\odot$ yr$^{-1}$.

We compare these results to those inferred using standard emission line diagnostics. 
Assuming H$\alpha/$H$\beta=2.86$ \citep{Osterbrock1989}, we derive a SFR(H$\alpha$)$= 1.2\pm 0.2 M_\odot \textrm{yr}^{-1}$
\citep{Kennicutt1998}.
This is consistent with the SFR derived using the [OII] line luminosity,  SFR([OII])$= 1.7\pm0.7 M_\odot \textrm{yr}^{-1}$ \citep{Kennicutt1998}, 
and only slightly lower than the SFR from SED modeling, suggesting
that most of the star formation activity is not obscured by dust. 
Overall, these results are
in keeping with the properties of sGRB host galaxies. 



\subsection{GRB 200522A}
\label{sec: 200522A section}

\subsubsection{Afterglow Properties}
\label{sec: 200522A_afterglow_overview}

Before introducing our broadband modeling, we start with some basic considerations on the afterglow properties. 
In X-rays, the observed spectral index $\beta_X$=0.45$\pm$0.18 suggests
that $\nu_m<\nu_\textrm{X}<\nu_c$, otherwise $p$\,$\lesssim$1.5. Here, $\nu_m$ is the injection frequency of electrons, and $\nu_c$ is the cooling frequency.
A consistent spectral index (at $\sim$\,$2\sigma$) is derived from the optical and nIR observations, $\beta_{OIR}$=1.1$\pm$0.3, suggesting the optical/nIR and X-ray data could lie on the same spectral segment. We therefore fit the broadband SED (X-ray/optical/nIR) at $T_0+3.5$ d 
with an absorbed power-law model \texttt{redden$*$}\texttt{tbabs(zdust(powerlaw))}\footnote{The command \texttt{redden} corrects optical/nIR wavelengths for extinction within the Milky Way \citep{Cardelli1989}, whereas  \texttt{zdust} accounts for intrinsic dust extinction within the host galaxy \citep{Pei1992}. We selected the \texttt{method} that uses Milky Way extinction curves.}
within \texttt{XSPEC}.  We fix the Galactic column density $N_H$=2.9$\times 10^{20}$ cm$^{−2}$, and $R_V=A_V/E(B-V)$=3.1 \citep{Rieke1985}. As there is no X-ray observation at this time, we re-scale the flux of the early-time XRT spectrum to the predicted value at 3.5 d. This fit yields $\beta_\textrm{OX}$=0.77$\pm$0.05 and $E(B-V)$=0.12$^{+0.12}_{-0.08}$~mag ($\chi^2$=71 for 78 dof), consistent with $\nu_m<\nu_\textrm{IR}<\nu_\textrm{X}<\nu_c$ and $p=2.54\pm0.10$. 
However, for this value the predicted temporal slope, $\alpha$=3$\beta$/2$\approx$1.16, is steeper than the slope, 
$\alpha_X$=0.84$\pm$0.04, measured in X-rays.

In GRB afterglows, it is not uncommon to observe a shallow temporal decay, not accounted for by the simple FS model. In general, this behavior is observed in the early ($<$1000 s) lightcurve, whereas later observations tend to be consistent with standard closure relations. Indeed, in this case too, we find that, by excluding the early X-ray data, 
the temporal slope steepens to $\alpha_X$=1.04$\pm$0.08 in agreement with the FS predictions. Therefore, in our modeling we only consider the late ($>$1000 s) X-ray data, as earlier epochs could be affected by complex factors (e.g., energy injection, jet structure) not included in the basic FS model. 

In the radio band, the afterglow is detected at a nearly constant level in the two early epochs (at $T_0+0.23$ and $T_0+2.2$ d), and then seen to fade \citep[see][]{Fong2020}. 
In the simple FS model, the radio emission is expected to either rise as $\nu^{1/3}$, if below the spectral peak ($\nu\!<\!\nu_m$), or decay as $\nu^{-(p-1)/2}$ when above the peak ($\nu\!>\!\nu_m$). 
The observed flattening can therefore be explained only if the synchrotron peak crosses the radio band between 0.1 and 2 days. 
Alternatively, the flat radio lightcurve could reveal the presence of a 
reverse shock (RS) component contributing at early times \citep{Fong2020}, as observed in other bright short GRBs \citep[e.g.,][]{Soderberg2006, Troja2019,Lamb2019,Becerra19}. Both scenarios are considered in our modeling.

We include an additional systematic uncertainty in the radio data to account for the scattering and scintillation of radio wavelengths due to interstellar scintillation (ISS). We adopt the `NE2001' model \citep{Cordes2002}, which yields a scattering measure $\textrm{SM}=1.9\times10^{-4}$ kpc m$^{-20/3}$ and a transition frequency $\nu_0\approx8$ GHz for the direction of GRB 200522A. Radio observations at $\nu_R<\nu_0$ can be effected by strong scattering when the angular extent of the GRB jet is $\theta_\textrm{GRB}<\theta_{F0}\!\approx\! 1\, \micro$as. In our afterglow modeling, we find that this condition, $\theta_\textrm{GRB}<\theta_{F0}$, is satisfied at $\lesssim 2.5$ d. Therefore, we include a 30\% systematic uncertainty, added in quadrature with the statistical uncertainty, to account for the effects of ISS. 

\begingroup
\renewcommand{\arraystretch}{1.5}
\begin{table*}
         \caption{The fit results of our afterglow modeling for GRB 200522A. The median and 68\% confidence interval of the marginalized posterior probability for each parameter corresponding to each mode determined from our MCMC fitting for the FS+Ext, Gauss+FS+Ext, FS+BB, and FS+BB+Ext models (see \S \ref{sec: 200522A_afterglow_overview}). 
         The FS+Ext model has only one solution mode, whereas the Gauss+FS+Ext model has two modes which are degenerate.
         The three modes for the FS+BB and FS+BB+Ext models are: a solution with a late radio peak at $\gtrsim\!10$ d (Mode I), a radio peak between 2-5 d that fits the radio detection at $2$ d (Mode II), and an early radio peak at 0.3-0.8 d which describes both VLA radio detections at $0.2$ and $2$ d (Mode III). Row 15 shows the minimum $\chi^2$ value associated to each mode.
         Rows 16-17 display the WAIC score of the expected log predictive density (\emph{elpd}), and the WAIC score difference, $\Delta$WAIC$_{\mathrm{elpd}}$, compared to the FS+Ext model. The WAIC score is computed for the overall model fit (see Table \ref{tab: fit_results}, available in the online version), and not for each individual mode.
         }
        \begin{tabular}{lrrrrrrrrrr}
        \hline
        \hline
        \multirow{2}{*}{Parameter} & FS + Ext & \multicolumn{2}{c}{Gauss + FS + Ext} & \multicolumn{3}{c}{FS + BB} &  \multicolumn{3}{c}{FS + BB + Ext}  \\
        \cmidrule(lr){2-2} \cmidrule(lr){3-4} \cmidrule(lr){5-7}  \cmidrule(lr){8-10} 
     & Mode I$^{a}$ & Mode I & Mode II$^{b}$ & Mode I & Mode II & Mode III$^{c}$ & Mode I & Mode II & Mode III \\
         \hline
  $\log_{10} E_0$ (erg)    
    & $50.69^{+0.08}_{-0.07}$
  &  $53.2^{+0.8}_{-0.7}$
  &  $51.2^{+1.0}_{-0.5}$
  &  $52.0^{+1.0}_{-0.8}$
  &  $51.7^{+1.2}_{-0.6}$
  & $51.9^{+1.0}_{-0.6}$
  &  $52.0^{+1.0}_{-0.8}$
  & $51.7^{+1.2}_{-0.6}$
  & $52.1^{+1.0}_{-0.7}$
\\
  $\theta_c$ (rad)      
    & $0.16^{+0.04}_{-0.03}$
  & $0.38^{+0.26}_{-0.13}$
  & $0.14^{+0.44}_{-0.11}$
  & $0.32^{+0.32}_{-0.23}$
  & $0.10^{+0.10}_{-0.06}$
  & $0.11^{+0.14}_{-0.07}$
  &  $0.31^{+0.32}_{-0.23}$
  & $0.10^{+0.10}_{-0.06}$
  & $0.09^{+0.12}_{-0.05}$
\\
  $\theta_v$ (rad)   
    & --
 & $0.60^{+0.13}_{-0.20}$
  & $0.10^{+0.17}_{-0.05}$
  & --
  & --
  & --
  & --
  & --
  & --
\\
  $\theta_w$ (rad)  
    & --
  & $1.04^{+0.36}_{-0.33}$
  & $0.36^{+0.62}_{-0.20}$
  & --
  & --
  & --
  & --
  & --
  & --
\\
  $\log_{10} E_j$  (erg) 
    &  $48.8^{+0.2}_{-0.1}$
  & $52.3^{+0.6}_{-0.6}$
  & $49.3^{+0.8}_{-0.3}$
  &  $50.5^{+1.2}_{-1.0}$
  &  $49.5^{+0.6}_{-0.4}$
  & $49.7^{+1.1}_{-0.8}$
  & $50.5^{+1.3}_{-1.0}$
  & $49.5^{+0.5}_{-0.4}$
  & $49.7^{+1.1}_{-0.9}$
\\
  $\log_{10} n_0$  (cm$^{-3}$) 
    & $-2.7^{+0.3}_{-0.2}$
  &  $-5.5^{+0.6}_{-0.3}$
  & $-2.8^{+0.3}_{-0.7}$
  & $-4.7^{+1.2}_{-0.9}$
  & $-3.1^{+1.3}_{-1.5}$
  & $-2.2^{+1.7}_{-1.9}$
  & $-4.7^{+1.2}_{-0.9}$
  & $-3.2^{+1.3}_{-1.6}$
  & $-2.4^{+1.6}_{-1.7}$
 \\
  $p$        
    & $2.32^{+0.19}_{-0.10}$
  & $2.93^{+0.19}_{-0.11}$
  & $2.51^{+0.23}_{-0.28}$
  & $2.50^{+0.12}_{-0.18}$
  & $2.13^{+0.08}_{-0.07}$
  &  $2.04^{+0.04}_{-0.02}$
  & $2.49^{+0.12}_{-0.19}$
  & $2.14^{+0.08}_{-0.07}$
  & $2.04^{+0.04}_{-0.02}$
\\
  $\log_{10}\varepsilon_e$  
    &  $-0.52^{+0.03}_{-0.05}$
  & $-0.62^{+0.11}_{-0.16}$
  & $-0.53^{+0.04}_{-0.10}$
  & $-0.66^{+0.13}_{-0.20}$
  & $-0.68^{+0.15}_{-0.26}$
  & $-0.68^{+0.15}_{-0.30}$
  & $-0.66^{+0.13}_{-0.20}$
  & $-0.68^{+0.15}_{-0.25}$
  & $-0.71^{+0.17}_{-0.32}$
\\
  $\log_{10}\varepsilon_B$    
    & $-0.57^{+0.07}_{-0.13}$
  & ${-2.1}^{+1.0}_{-1.1}$
  & $-0.7^{+0.2}_{-0.8}$
  & $-1.9^{+1.0}_{-1.5}$
  & $-2.3^{+1.1}_{-1.5}$
  &  $-3.1^{+1.3}_{-1.2}$
  & $-1.9^{+1.0}_{-1.5}$
  & $-2.4^{+1.1}_{-1.4}$
  & $-3.2^{+1.2}_{-1.2}$
 \\
       $E(B-V)$  (mag)
 &  $0.16^{+0.08}_{-0.07}$
  & $0.08^{+0.06}_{-0.04}$
  & $0.18^{+0.10}_{-0.08}$
  & --
  & --
  & -- 
  & $0.18^{+0.21}_{-0.13}$
  &$0.23^{+0.22}_{-0.16}$
  & $0.21^{+0.18}_{-0.14}$
\\
\hline
 $R$ ($10^{15}$ cm) 
    & --
    & --
  & --
  & $2.22^{+1.13}_{-0.73}$
  & $2.28^{+0.93}_{-0.61}$
  & $2.16^{+0.76}_{-0.51}$
  & $1.61^{+0.81}_{-0.39}$
  &$1.72^{+0.64}_{-0.36}$
  & $1.74^{+0.55}_{-0.34}$
\\
 $T$ (K)
    & --
    & --
  & --
  & $4050^{+950}_{-700}$
  & $4300^{+800}_{-700}$
  & $4600^{+800}_{-700}$
  & $5600^{+1600}_{-1500}$
  & $5800^{+1400}_{-1300}$
  & $5850^{+1200}_{-1100}$
\\
  \hline
  \hline
  $\chi^2$
  & 10.5
  & 8.8
  & 5.4
  & 5.8
  & 3.8
  & 3.9
  & 6.0
  & 4.1
  & 4.3
  \\
    \hline
  WAIC$_{\mathrm{elpd}}$      
  &  $150\pm19$  
  &  \multicolumn{2}{c}{$88\pm19$}
  &  \multicolumn{3}{c}{$151\pm19$}
  &   \multicolumn{3}{c}{$154\pm18$}
 \\
  $\Delta$WAIC$_{\mathrm{elpd}}$
  &  --         
    &  \multicolumn{2}{c}{$-62\pm40$}
  &  \multicolumn{3}{c}{$1.2\pm3.2$}
  &  \multicolumn{3}{c}{$4.2\pm3.5$}
 \\ 
  \hline
    \end{tabular}
   \label{tab: fit_results_BB}
 \begin{flushleft}
     \quad \footnotesize{$^a$ This is the only mode of the solution for this model.} \\
    \quad \footnotesize{$^b$ This mode is comprised of two solutions, but due to the high degree of degeneracy it is not possible to further sort these solutions.} \\
        \quad \footnotesize{$^c$ This mode of the solution appears only when the first radio detection at $T_0+0.2$ d \citep{Fong2020} is included within the fit.}
\end{flushleft}
\end{table*}
\endgroup

Using an MCMC Bayesian fitting approach, outlined in \S \ref{sec: model_fitting_methods}, we explore four different models to describe the broadband afterglow of GRB 200522A from radio to X-rays. By assuming a top-hat jet, we consider (i) a forward shock with intrinsic extinction from the host galaxy (hereafter denoted by FS+Ext), (ii) a forward shock with an additional blackbody component (FS+BB, see \S \ref{sec: simple BB}), and (iii) a forward shock and simple blackbody with the addition of intrinsic extinction (FS+BB+Ext).  
For a structured jet, we consider (iv) a Gaussian profile, 
standard forward shock emission, and intrinsic extinction (Gauss+FS+Ext). The results are tabulated in Tables \ref{tab: fit_results_BB} and \ref{tab: fit_results} (available in the online version). We discuss the results of these fits for the models with only FS emission in \S \ref{sec: 200522A_afterglow_FS} and models with an additional BB component in \S \ref{sec: 200522A_afterglow_BB}. A comparison of the WAIC score difference between these models is presented in \S \ref{sec: waic_analysis}.
Lastly, in \S \ref{sec: 200522A_kilonova}, we explore the consistency of the predicted flux from our BB models with detailed kilonova simulations (\S \ref{sec: KN_grid}). 

\begin{figure} 
\centering
\includegraphics[width=\columnwidth]{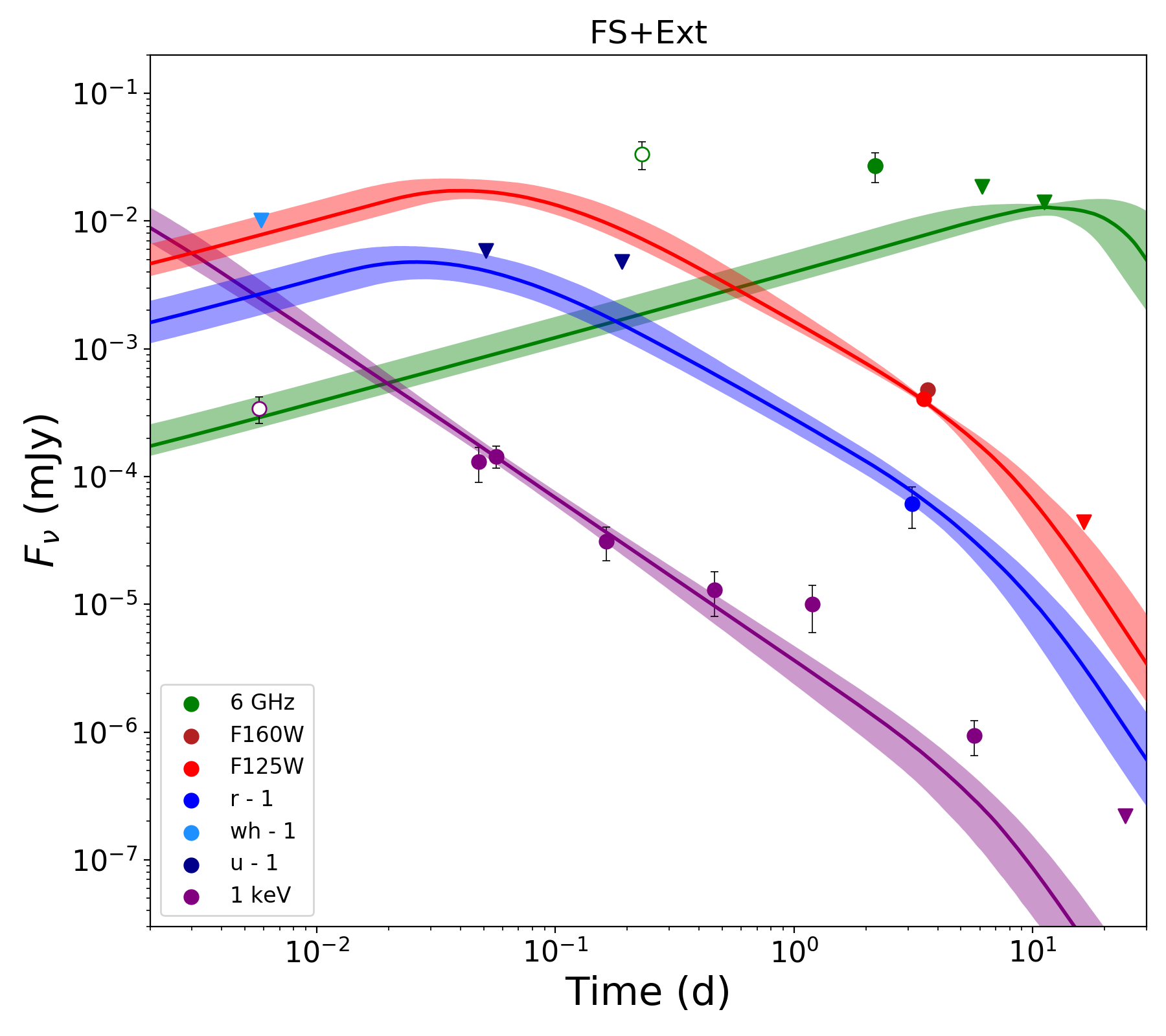}
\vspace{0.1cm}
\includegraphics[width=\columnwidth]{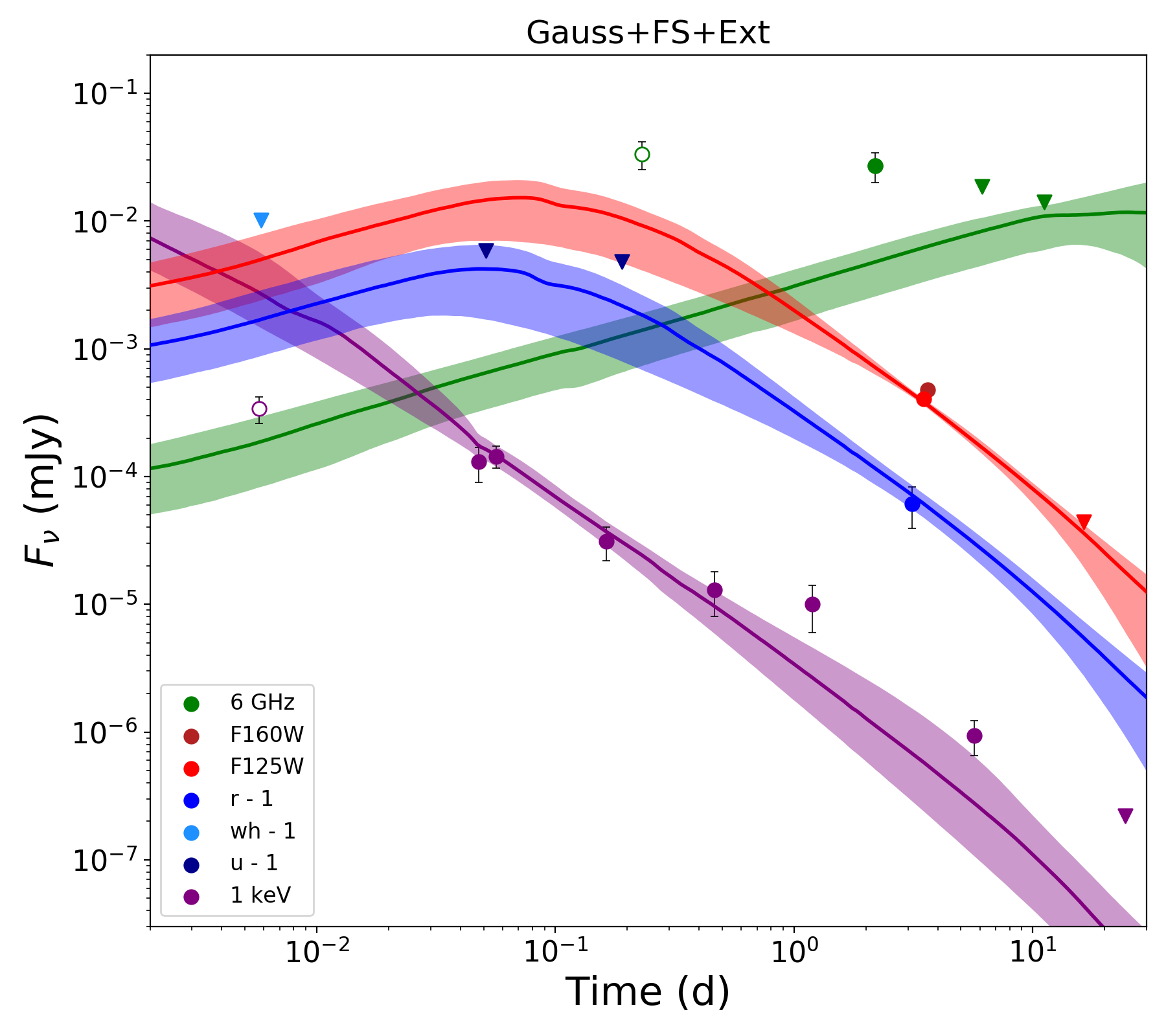}
\vspace{-0.7cm}
\caption{Broadband lightcurve of GRB 200522A compared to the standard forward shock with intrinsic extinction scenario for two angular profiles, tophat (Top: model FS+Ext) and Gaussian (Bottom: model Gauss+FS+Ext). The shaded regions mark the $1\sigma$ 
uncertainty in the model. Hollow circles mark data excluded from the fit (see \S \ref{sec: 200522A_afterglow_overview}). $3\sigma$ upper limits are denoted by downward triangles.}
\label{fig: 200522A_fs+ext}
\end{figure}

\begin{figure}
\centering
\includegraphics[width=\columnwidth]{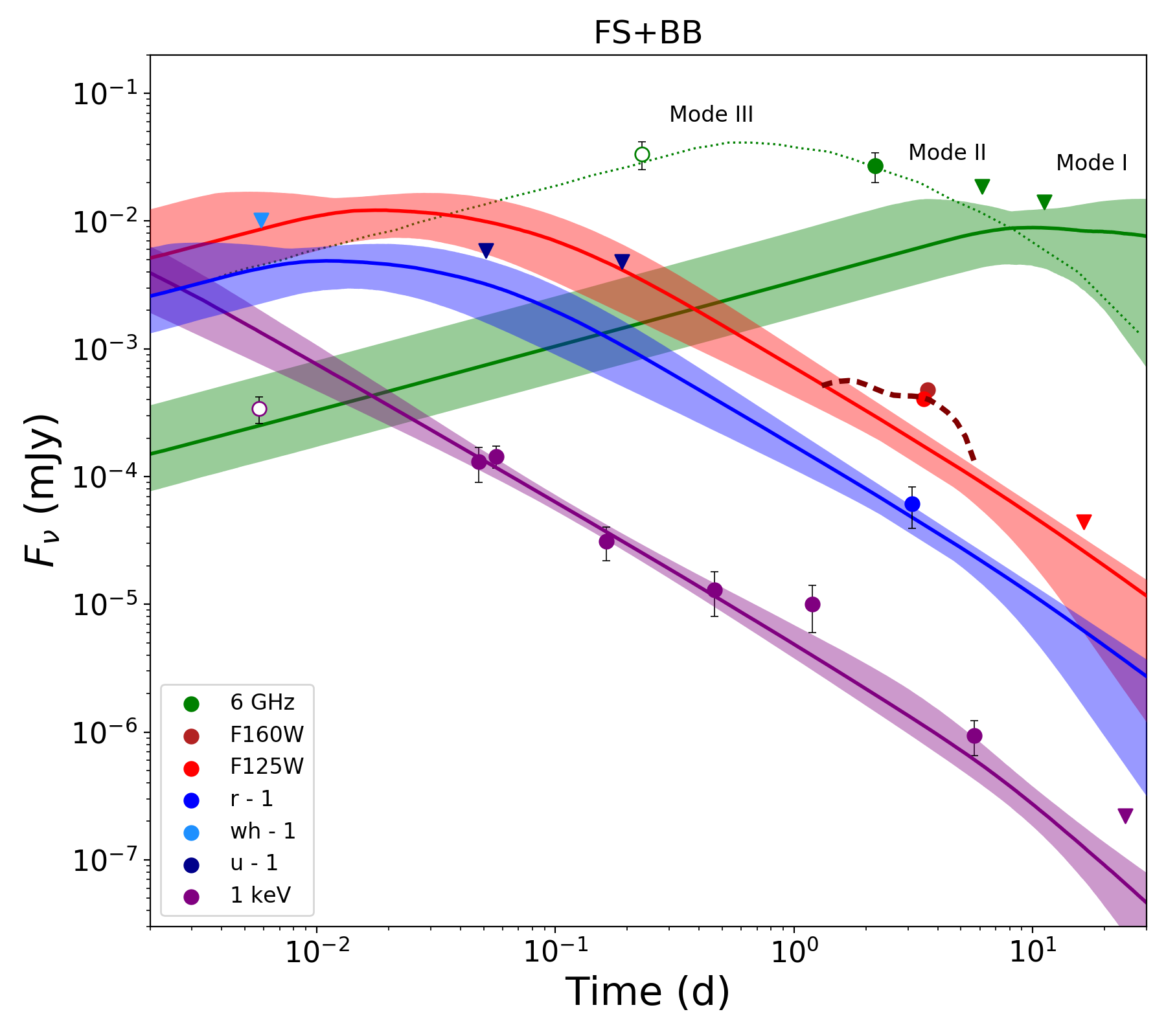}
\includegraphics[width=\columnwidth]{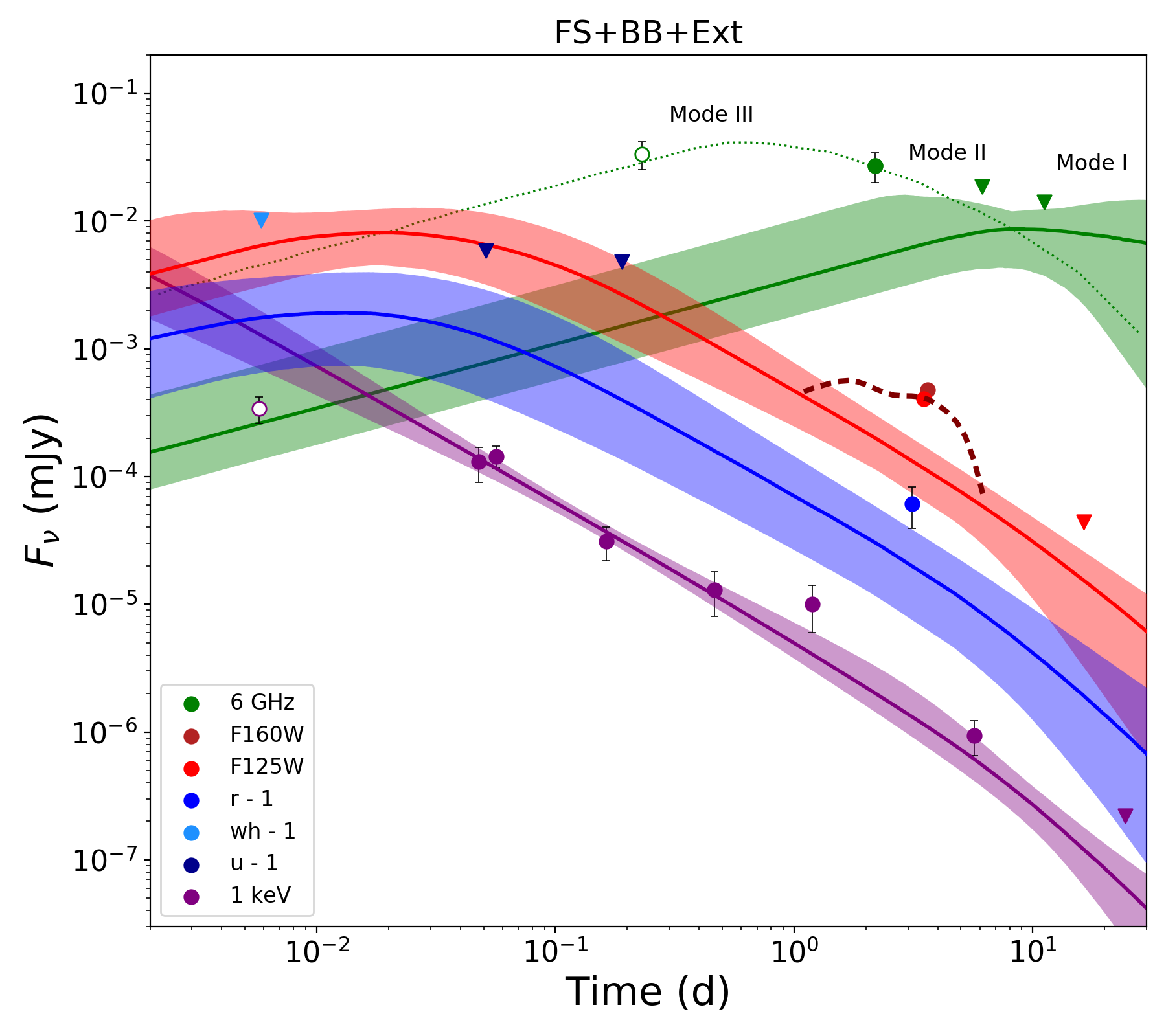}
\vspace{-0.5cm}
\caption{Same as in Figure~\ref{fig: 200522A_fs+ext}, but for the forward shock model with a blackbody component (Top: model FS+BB) and with intrinsic extinction (Bottom: model FS+BB+Ext). The shaded regions display only the afterglow contribution to the flux.
Multiple solutions are consistent with the data: a late radio peak at $\gtrsim\!10$ d (Mode I), a radio peak between 2-5 d (Mode II), and an early radio peak at 0.3-0.8 d (Mode III; dotted line). 
The excess emission (\textit{HST/F125W}) is compared to a simulated kilonova lightcurve (dashed maroon line) with properties: $M_{\textrm{ej,d}}$=$0.001M_\odot$, $v_{\textrm{ej,d}}$=$0.3c$, $M_{\textrm{ej,w}}$=$0.1\,M_\odot$, and $v_{\textrm{ej,w}}$=$0.15c$.
}
\label{fig: 200522A_fs+ext_BB}
\end{figure}

\subsubsection{Forward Shock Afterglow Models}
\label{sec: 200522A_afterglow_FS}

Here, we discuss the inferred parameters for the model with standard FS emission, including intrinsic extinction from the host galaxy, for two jet configurations (see \S \ref{sec: model_fitting_methods}): a top-hat jet viewed on-axis ($\theta_v\!\approx\!0$) and a Gaussian angular profile viewed at an arbitrary angle $\theta_v$. A comparison of these models to the broadband dataset of GRB 200522A is shown in Figure \ref{fig: 200522A_fs+ext}. In Supplementary Figures \ref{fig: 200522A_fs+ext_corner} and \ref{fig: 200522A_gauss+fs+ext_corner} we present the marginalized posterior distributions for each parameter from our MCMC fit for the tophat and Gaussian jet, respectively (these are available in the online version).
From this we identify that the Gaussian jet model has a bi-modal solution which we refer to as Mode I and Mode II; see Table \ref{tab: fit_results_BB}. 
Mode I yields an extremely large beaming-corrected energy $E_j\!\approx\!2\times 10^{52}$ erg and a very low circumburst density, $n_0\!\approx\!3\times 10^{-6}$ cm$^{-3}$, likely inconsistent with the observed offset \citep{OConnor2020}. Mode II instead yields more typical values $E_j\!\approx\!2\times 10^{49}$ erg, $n_0\!\approx\!2 \times 10^{-3}$ cm$^{-3}$,  consistent with those inferred for the top-hat jet model. 
Although both modes are statistically equivalent, we consider Mode I a less realistic description of the afterglow parameters. 

 Figure \ref{fig: 200522A_fs+ext} (top panel) shows that the standard FS model (FS+Ext) can describe the broadband dataset, except for the early radio and X-ray detections. An achromatic jet-break \citep{Rhoads1999,Sari1999} at $\approx$5~d is required by the data, which constrains the jet opening angle to $\theta_c$\,$\approx$\,0.16 rad (9$^{\circ}$). 
The Gaussian angular profile (bottom panel) leads to a slightly better description of the X-ray light curve. Due to its shallower temporal decline, it can more easily reproduce (within the $2\sigma$ uncertainty) the first X-ray detection at $<$1000 s
without requiring additional energy injection. 
However, unlike the top-hat jet model, the Gaussian angular profile does not provide a tight constraint on the jet's opening angle, $\theta_c$. Although the ratio $\theta_v/\theta_c$ is better constrained to $\theta_v/\theta_c \lesssim 2.3$.

Both these jet models underestimate the radio detections at $T_0+0.23$ d and at $T_0+2.2$ d. 
This could be attributed to synchrotron self Compton (SSC) and Klein-Nishina effects \citep[e.g., ][]{Jacovich2020}, not included in our code, and/or to a bright RS component.

\subsubsection{Afterglow Models including a Blackbody Component}
\label{sec: 200522A_afterglow_BB}

In this section, we present the multi-modal solutions for our afterglow models that include a simple blackbody component, with (FS+BB+Ext) and without extinction (FS+BB). The fit to the broadband dataset for each model is shown in Figure \ref{fig: 200522A_fs+ext_BB}, and the parameter values are presented in Supplementary Table \ref{tab: fit_results}. Marginalized posterior distributions for each parameter are displayed in Supplementary Figures \ref{fig: 200522A_fs+bb_corner} and \ref{fig: 200522A_fs+bb_corner2} for the FS+BB and FS+BB+Ext models, respectively (all supplementary figures and tables are available in the online version of the manuscript). Each model exhibits three modes, referred to as Mode I, Mode II, and Mode III, which are presented individually in Table \ref{tab: fit_results_BB}. 
Mode I is characterized by a late ($t>$10~d) radio peak and no jet-break,  
Mode II shows an earlier peak (2~d$<t<$5~d) and requires a jet-break, and 
Mode III\footnote{We note that Mode III appears with high significance only when the first radio detection is included in the fit.} also finds a jet-break and describes the first and second radio detections without a RS component due to an early radio peak (0.3-0.5~d). 
Although the MCMC algorithm cannot distinguish one of these modes as providing a better description of the data, we disfavor Mode I based on the extremely low circumburst density ($n\approx10^{-5}$\,cm$^{-3}$) and the phenomenology of sGRB afterglows. Such late radio peaks have not been observed in sGRBs, and Mode I is likely an artifact of the poorly constrained late-time radio dataset. 
We therefore favor either Mode II or Mode III as a more likely description of the jet parameters. Both of them constrain the jet-opening angle to $\theta_c$$\approx$\,0.10 rad (6$^{\circ}$). 

When including extinction, the only difference in the fit is in the temperature $T$ and radius $R$ of the blackbody component: 
the model without dust extinction requires a cooler thermal component (T$\approx$4,000~K) to match the optical data.
A higher temperature  (T$\gtrsim$6,500~K),
as predicted, e.g., in the magnetar-boosted model \citep{Fong2020}, tends to overpredict the optical
flux, unless dust extinction contributes to attenuate the observed emission.


\subsubsection{Model Comparison}
\label{sec: waic_analysis}
We perform a comparison of the models applied in this work using their WAIC scores, described in \S \ref{sec: model_fitting_methods} (see also \citealt{Troja2020,Cunningham2020}). We note that the WAIC analysis is not applicable to individual modes within the models, and that we only compare the overall model fits presented in Supplementary Table \ref{tab: fit_results}.

For the four models considered, the difference between the WAIC scores is not significant enough to statistically favor any of them. In particular, the addition of a blackbody to the FS fit is not required by the data. We find that the WAIC score of the FS+BB+Ext model only marginally (at the $1.2\sigma$ level) improves over the FS+Ext model. 

The most significant difference in WAIC score ($\Delta$WAIC$_{\mathrm{elpd}}=-62\pm40$) is between the FS+Ext and Gauss+FS+Ext models. A larger WAIC score implies a better description of the data (see \S \ref{sec: model_fitting_methods}), and, therefore, the FS+Ext model is marginally preferred at the $\approx1.4\sigma$ level. 

Our findings do not depend on the details of the data analysis, which yield slightly different magnitudes for the nIR counterpart (see \S \ref{sec: 200522A optical analysis}). In particular, we verified that, by using the values presented by \citet{Fong2020}, our results are unchanged. 
There is no significant difference between the posterior distributions of the fit parameters, and the WAIC score comparison continues to not favor any particular model fit.

\begin{figure} 
\centering
\includegraphics[width=\columnwidth, trim=50 25 50 80, clip]{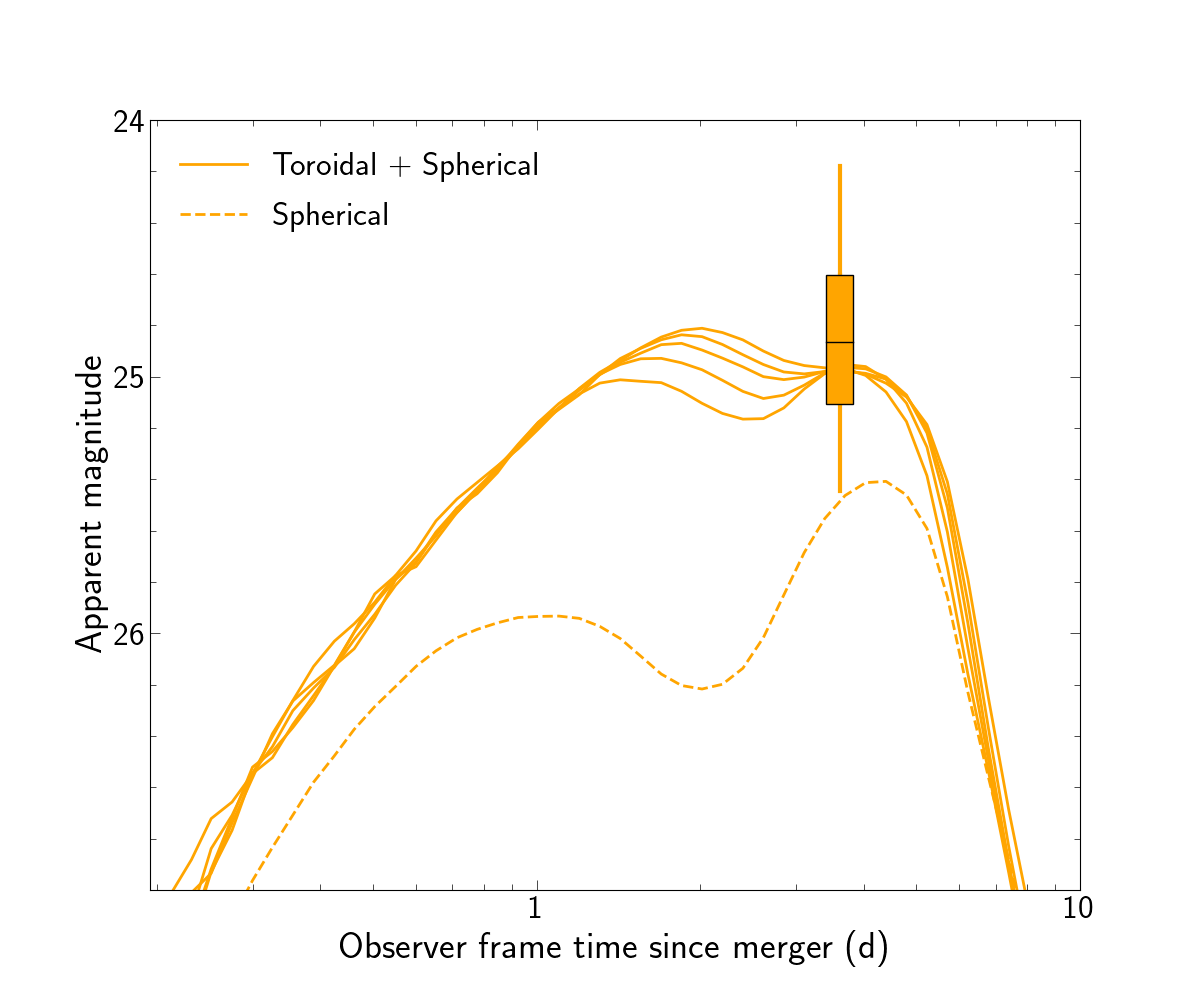}
\caption{Simulated kilonova lightcurves 
consistent with the blackbody flux posterior distribution in the \textit{HST}/$F160W$ filter. 
The box indicates the inner 50\% credible interval while the whisker spans the inner 90\% credible interval of the posterior distribution. The black line within the box corresponds to the median value. 
The solid lines correspond to on-axis emission from a two-component model with a spherical wind ejecta girdled by a toroidal belt of lanthanide-rich dynamical ejecta. 
For comparison, the dashed line shows a single-component spherical morphology with identical properties of the wind ejecta ($M_{\textrm{ej}}$\,=0.1\,$M_\odot$,
$v_{\rm ej,w}$=0.15$c$, $Y_e$=0.27). 
}
\label{fig: 200522A_kilonova_constraints}
\end{figure}

\subsubsection{Kilonova Constraints}
\label{sec: 200522A_kilonova}

 We use the simple blackbody component, described in \S \ref{sec: kilonova_models}, to constrain the contribution of a possible kilonova to the observed nIR emission. The blackbody luminosity lies in the range $L_\textrm{F125W}\approx(7-19)\times 10^{41}$ erg s$^{-1}$ and $L_\textrm{F160W}\approx(7-15)\times 10^{41}$ erg s$^{-1}$ (observer frame). Whereas the constraint on a thermal contribution at (observer frame) optical wavelengths has a greater uncertainty $L_r\approx (2-22) \times 10^{41}$ erg s$^{-1}$. These values are somewhat larger than observed from other candidate kilonovae and AT2017gfo, which are $\approx$\,(1-3)$\times 10^{41}$ erg s$^{-1}$ at similar times.

 We utilize our MCMC algorithm to determine a posterior distribution on the temperature $T$ and radius $R$ of the blackbody (Table \ref{tab: fit_results_BB}), from which we compute a distribution for the emitted flux from the blackbody in each filter. The flux posterior distributions are then compared to the LANL suite of kilonova simulations (see \S \ref{sec: KN_grid}) in order to estimate the ejecta masses and velocities required to reproduce the observations.
 Figure~\ref{fig: 200522A_kilonova_constraints} presents five kilonova lightcurves (solid lines) consistent with the black body flux estimated in the FS+BB+Ext model. 
 They lie within the inner 50\% credible interval of both \textit{HST}/$F125W$ and \textit{HST}/$F160W$ posteriors. 
 As the FS contribution likely dominates at optical wavelengths, we do not require consistent lightcurves to reside in the 50\% credible interval of the Gemini/\textit{r}-band constraint. 

These results indicate that any thermal component, as constrained from our broadband modeling, agrees with a radioactively powered kilonova emission. 
 The five consistent lightcurves span a wide range of dynamical ejecta masses (0.001\,$M_\odot$\,$\lesssim$\,$M_{\textrm{ej,d}}$\,$\lesssim$0.1\,$M_\odot$), but share many other properties, including a wind ejecta mass of $M_{\textrm{ej,w}}$\,=0.1\,$M_\odot$, wind ejecta velocity of $v_{\textrm{ej,w}}$=0.15$c$, dynamical ejecta velocity of $v_{\textrm{ej,d}}$=0.3$c$, and spherical wind morphology. 
 These observations provide stronger constraints on the wind ejecta mass as the $F125W$ and $F160W$ filters probe rest frame optical wavelengths.
We emphasize  
that these 5 lightcurves represent only a small subset of kilonova parameters capable of reproducing the flux posteriors. 

This result differs from the findings of \citet{Fong2020}, who argue for 
an additional power source (e.g., an enhanced radioactive heating rate or a magnetar) to boost the kilonova luminosity to values higher than AT2017gfo. 
This difference arises despite comparable predictions for the nIR thermal emission, with their predicted luminosities ($L_\textrm{F125W}\approx(9.5-12.3)\times 10^{41}$ erg s$^{-1}$ and $L_\textrm{F160W}\approx(8.9-11.4)\times 10^{41}$ erg s$^{-1}$) fully encapsulated within our luminosity posterior distribution.
Our multi-dimensional kilonova models can reproduce this 
range of values by incorporating the same physics adopted to model AT2017gfo \citep{Troja2017,Evans2017,Tanvir2017,Wollaeger2018}, 
including a thermalizable heating rate of $\approx10^{10}$ erg g$^{-1} $s$^{-1}$ at $t=1$~day. 
In addition, they capture the multi-component character of kilonova ejecta, leading to an enhancement of the emission via photon reprocessing \citep{Kawaguchi2020}.
This is illustrated in Figure~\ref{fig: 200522A_kilonova_constraints}, 
which compares our models (solid lines), produced by a spherical wind component girdled with a toroidal lanthanide-rich ejecta, with the emission produced by a single-component spherical morphology (dashed line), with properties ($M_{\textrm{ej}}$\,=0.1\,$M_\odot$, $v_{\rm ej,w}$=0.15$c$, $Y_e$=0.27) 
 similar to the models used in \citet{Fong2020}.
The addition of the toroidal belt produces a $\approx$\,1~mag enhancement of the \textit{HST/F160W} flux compared to the single-component model. 
This is attributed to the reprocessing of photons emitted from the 
spherical wind by the high-opacity ejecta. 
Photons absorbed by the toroidal component preferentially diffuse towards the polar regions, and are re-emitted at redder wavelengths. The flux enhancement at optical wavelengths is instead negligible. 
This effect is more prominent in events viewed 
along the polar axis, such as GRB~200522A.

\subsubsection{Environment}
\label{sec: 200522A_host}


GRB 200522A is located within a bright galaxy, SDSS J002243.71-001657.5, at $z\!=\!0.554$. The projected offset from the galaxy's nucleus is $R=0.16\pm0.04\arcsec$ ($1.07 \pm 0.27$ kpc);
in the bottom 15\% of the sGRB offset distribution \citep{FongBerger2013}.
Using \texttt{GALFIT}, we derive a projected half-light radius 
$R_e=0.60 \pm 0.02\arcsec$, corresponding to 4.0 $\pm$0.1 kpc, and a normalized offset of $R/R_e\approx 0.27\pm0.07$. 
The chance alignment between the GRB and the galaxy is small:
we determine $P_{cc}$=0.002 using the r-band magnitude, and $P_{cc}$=0.003 using the $F160W$ filter. Both the GRB offset and the galaxy's angular size
contribute to determine this value \citep{Bloom2002}.
As shown in Figure~\ref{fig: 200522A_RGB}, there are two nearby red galaxies seen at projected angular offsets of $2.5\arcsec$ and $3.9\arcsec$ with $F160W$\,$\approx$\,23.5 and 20.8 AB mag, respectively. 
The probability of chance coincidence is $P_{cc}$=0.11 and $0.05$, respectively. Therefore, we consider SDSS J002243.71-001657.5 to be the likely host galaxy of GRB~200522A. 

\begin{figure} 
\centering
\includegraphics[width=\columnwidth]{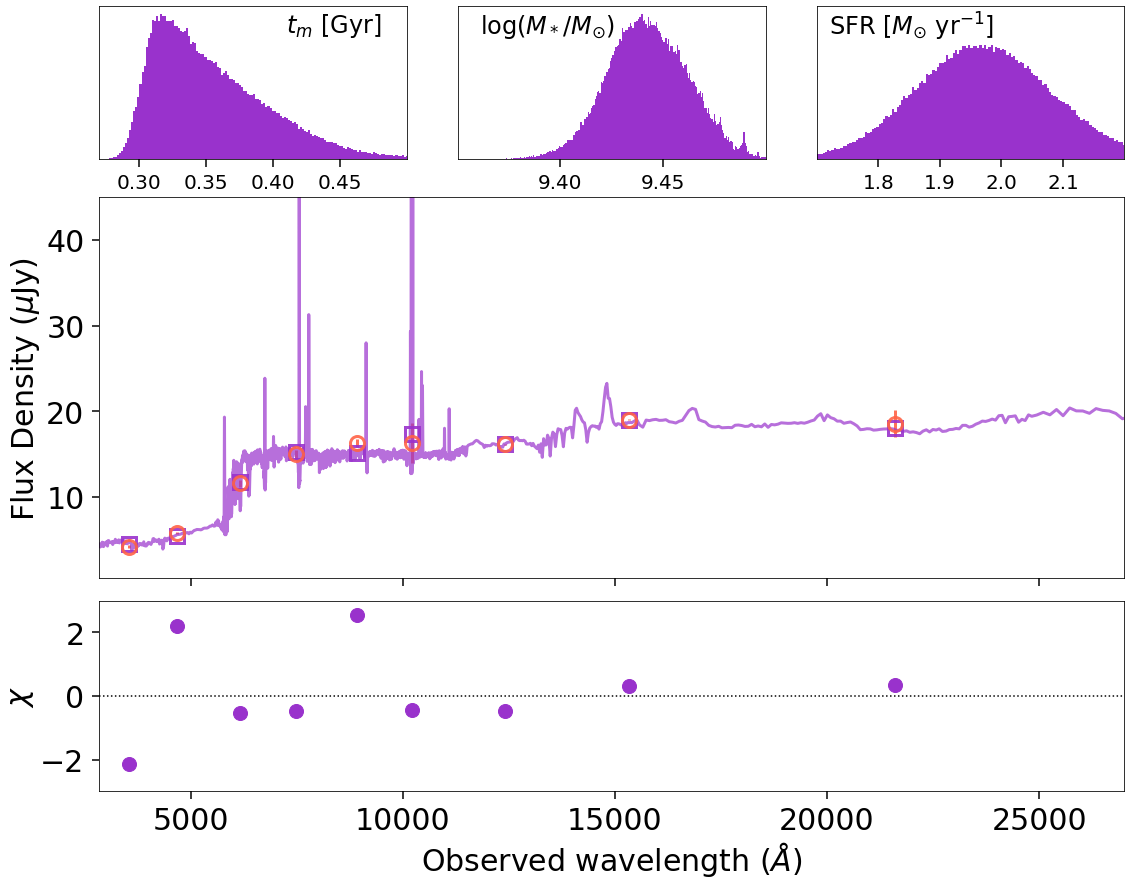}
\vspace{-0.5cm}
\caption{The best fit model spectrum (solid line) and photometry (squares) characterizing the host galaxy SED for GRB 200522A. The observed photometry (circles), corrected for Galactic extinction, is also shown. The bottom panel shows the fit residuals. The stamps (top) demonstrate posterior distributions for $t_m$, $\log(M_*/M_\odot)$, and the SFR.}
\label{fig: 200522A_SED}
\end{figure}

We determine the physical properties of the galaxy using the methods described in \S \ref{sec: SED fitting}. The resulting best fit model is shown in Figure \ref{fig: 200522A_SED}. We find an integrated extinction $E(B-V)=0.02\pm0.01$ mag, a metallicity $Z_*/Z_\odot=1.35^{+0.16}_{-0.25}$, an $e$-folding time $\tau=0.13^{+0.07}_{-0.03}$ Gyr, a stellar mass $\log(M_*/M_\odot)=9.44\pm0.02$, a young stellar population $t_m=0.35^{+0.08}_{-0.04}$ Gyr, and a star-formation rate SFR $=2.00\pm0.12$ $M_\odot$ yr$^{-1}$. These values are broadly consistent with those inferred by \citet{Fong2020}, although they derive a slightly higher stellar mass. The stellar mass and age derived here are on the low end of the distributions for sGRB host galaxies, but are not unique to the population \citep{Leibler2010}.

Moreover, we perform standard emission line diagnostics on the spectrum described in \S \ref{sec: 200522A_spectroscopy}. The line flux ratio H$\gamma/$H$\beta=0.25\pm0.11$ does not provide a strong constraint on the intrinsic extinction $E(B-V)$. 
\citet{Fong2020} infer negligible extinction, however, based on their reported line ratio H$\alpha/$H$\beta=2.9\pm0.9$, we derive $E(B-V)\lesssim 1.3$ mag  ($3\sigma$) which does not rule out the presence of moderate extinction along the GRB line of sight. 
The dominant emission lines imply significant on-going star formation. 
We derive a SFR(H$\alpha$)\,$\approx5.4\pm1.4$\,$M_\odot \textrm{yr}^{-1}$ under the assumption H$\alpha/$H$\beta=2.86$, and SFR([OII])$=7.3\pm1.9$ \,$M_\odot \textrm{yr}^{-1}$ using the [OII] line luminosity \citep{Kennicutt1998}.


\section{Conclusions}
\label{sec: conclusions}

We have presented an analysis of the multi-wavelength datasets for two sGRBs at $z\sim0.5$: GRB 160624A at $z=0.483$ and GRB 200522A at $z=0.554$. These two events demonstrate the wide range of diversity displayed by sGRBs in terms of both their emission and environment. We utilize the broadband datasets for these two events to constrain the presence of kilonova emission arising after the compact binary merger.

Gemini and  \textit{HST} observations of GRB~160624A place some of the deepest limits on both optical and nIR emission from a short GRB. For this event, we find that the bright short-lived X-ray counterpart is likely related to long-lasting central engine activity,  whereas emission from the forward shock is not detected.
Thanks to the negligible afterglow contribution, 
we can robustly constrain emission from a possible kilonova. By comparing our limits to a large suite of detailed simulations,  we derive a total ejecta mass $\lesssim 0.1 M_\odot$, favoring wind ejecta masses $M_\textrm{ej,w}\lesssim 0.03 M_\odot$. Any kilonova brighter than AT2017gfo ($\sim$30\% of the simulated sample) can be excluded by our observations. 

A late-time \textit{Chandra} observation places a deep upper limit on the X-ray emission ($L_X\lesssim 1.1\times10^{42}$ erg/s), which we use to constrain the presence of a long-lived magnetar. Radiation from the magnetar nebula could ionize the low-opacity wind ejecta on a timescale of a few days/weeks \citep{Metzger2014}, allowing for X-rays from the central engine to escape. The lack of X-ray detection at $T_0+5.9$ d (rest frame) implies a magnetar with a strong magnetic field ($\gtrsim\!10^{15}$ G) and a large mass of ejecta ($\gtrsim\!10^{-2} M_\odot$). However, this amount of ejecta, when re-energized by the magnetar's spin-down emission, should also produce a luminous optical and infrared signal detectable with both Gemini and \textit{HST}. Therefore, our multi-wavelength dataset disfavors a stable magnetar remnant for GRB 160624A.

Very different is the phenomenology of GRB 200522A, which has a long-lived X-ray emission, largely consistent 
with standard forward shock models, and a bright nIR counterpart. 
Its unusual red color ($r-H\!\approx\!1.3$) at a rest frame time $\sim 2.3$ d
is suggestive of a kilonova, although the inferred luminosity of  $L_\textrm{F125W}\approx(7-19)\times 10^{41}$ erg s$^{-1}$ is above the typical range observed in other candidate kilonovae \citep{Tanvir2013,Yang2015,Troja2019,Ascenzi2019kn}. 
The identification of a kilonova is complicated by the bright and long-lived afterglow contribution, as well as by the limited color information available for this event. 
Our thorough modeling of the multi-wavelength dataset finds that 
a standard forward shock model represents a good description of the X-ray, optical and nIR dataset provided that a modest amount of extinction, $E(B-V)$\,$\sim$0.1-0.2 mag, is present along the GRB sightline. This value is not at odds with the negligible 
extinction from galaxy's SED modeling, which is integrated over the whole galaxy and may not be representative of the GRB sightline, 
as found in other short GRBs
\citep[e.g., GRB 111117A;][]{Sakamoto2013,Selsing2018}.
A moderate extinction is consistent with the constraints from optical spectroscopy, $E(B-V)$\,$\lesssim$1.3 mag. From X-ray spectroscopy we derive a similar upper bound of $E(B-V)$\,$\lesssim$1.1 mag, assuming a galactic dust-to-gas ratio \citep{Guver2009}.
The location of GRB 200522A within its host ($\sim 1.1$ kpc from the center) and evidence for ongoing star-formation ($\sim$2-6 $M_\odot$ yr$^{-1}$) also support that dust effects might not be completely negligible, as instead assumed by \citealt{Fong2020}. 

We constrain the afterglow parameters to: $E_j\approx 6 \times 10^{48}$ erg, $n_0\approx 2 \times 10^{-3}$ cm$^{-3}$, $p\approx 2.3$, and identify the presence of a jet-break at $\approx$\,5~d. From this we derive an opening angle of $\theta_c\approx 0.16$ rad (9$^{\circ}$),  providing additional evidence that short GRB outflows are collimated into jets with a narrow core \citep[e.g.][]{Burrows2006,Troja2016}. 
Since this GRB was likely observed close to its jet-axis ($\theta_v/\theta_c\lesssim$2; for comparison, GW170817 had $\theta_v/\theta_c\approx 6$; \citealt{Ryan19,Beniamini20}), no significant constraint can be placed on the jet structure and a simple top-hat jet profile seen on-axis ($\theta_v\approx 0$) already provides an adequate description.
Our best fit model cannot reproduce the early radio detection with a simple forward shock emission, suggesting the presence of an early reverse shock component (see also \citet{Jacovich2020} for a discussion of SSC effects).

A different interpretation is of course that the red color of the optical/nIR emission marks the onset of a
luminous kilonova. Statistically, this scenario (models FS+BB and FS+BB+Ext in Table~\ref{tab: fit_results}) is not preferred over a simple afterglow model (FS+Ext in Table~\ref{tab: fit_results}) and therefore a kilonova component is not required by the data. 
Nevertheless, we explore the range of kilonova models consistent with our dataset. 
A radioactively-powered kilonova with wind ejecta mass $M_\textrm{ej,w}=0.1 M_\odot$, wind velocity $v_\textrm{ej,w}=0.15c$, and dynamical ejecta velocity $v_\textrm{ej,d}=0.3c$ is capable of reproducing the observed nIR emission
when considering a toroidal morphology for the lanthanide-rich ejecta \citep{Korobkin2020}. 
This geometry can naturally produce brighter transients than spherically symmetric counterparts of equal mass, expansion velocity, and radioactive heating. 
The range of dynamical ejecta masses is however not well constrained, as the rest frame wavelengths probed by the observations lie in the optical band.

The ejecta mass implied by the nIR luminosity is slightly larger than the values derived for AT2017gfo ($M_{\textrm{ej,w}}$\,$\approx$0.02-0.07\,$M_{\odot}$), and pushes the boundaries of a standard NS merger model. Numerical simulations of accretion discs indicate that, following a NS merger, $\approx$10--40\% of the disc can become unbound and form a massive outflow along the polar axis \citep[see, e.g.,][]{Perego2017}. 
Since these mergers can form discs up to $\approx$0.3\,$M_{\odot}$, a wind ejecta mass $M_\textrm{ej,w}=0.1 M_\odot$ is still within the range of possible outcomes \citep[e.g.,][]{Giacomazzo2013b,Radice2018c,Kiuchi2019,Fernandez2020,Fujibayashi2020}. Such a large amount of wind ejecta can be expected for a progenitor system comprising either low-mass NSs (for a NS-NS binary) or a NS-BH system where the BH has a favorable combination of high spin, $\chi\!>\!0.5$, and low mass, $M_\textrm{BH}<5 M_\odot$ \citep[e.g.,][]{Giacomazzo2013b,Fernandez2020}.


Moreover, our derived ejecta mass is limited by the resolution of the LANL simulations which do not fully probe the range 0.03-0.1\,$M_\odot$. It is conceivable that a finer sampled grid of simulated light curves could find additional solutions in this mass range. 
Based on these considerations, we find no compelling evidence for a magnetar-powered kilonova discussed by \citealt{Fong2020}. 

Our study of GRB~160624A and GRB~200522A demonstrates that deep \textit{HST} observations can probe an interesting range of kilonova behaviors out to $z$\,$\sim$\,0.5. 
However, whereas sensitive  \textit{HST} imaging can detect the bright kilonova emission, it is not sufficient to unambiguously identify a kilonova and disentangle it from the standard afterglow. 
Observations of GRB~200522A, based on a single multi-color epoch, can not break the degeneracy between the different models and, overall, the presence of a kilonova can not be confidently established. As shown by previous cases, such as GRB~130603B and GRB~160821B, multi-epoch multi-color observations are essential
for the identification of a kilonova bump. 
Moreover, at these distances, we found that the component of lanthanide-rich ejecta is only weakly constrained by the  \textit{HST} observations, with an allowed range of masses that spans two orders of magnitude (0.001-0.1 $M_\odot$). Future IR observations with \textit{JWST} will be pivotal to constrain the properties of lanthanide-rich outflows from compact binary mergers.

\section*{Acknowledgements}
The authors would like to acknowledge the anonymous referee for useful comments that improved the manuscript.
B.O. acknowledges Paz Beniamini and Amy Lien for useful discussions. 
B.O., E.T. and S.D. were supported in part by the National Aeronautics and Space Administration through grants NNX16AB66G, NNX17AB18G, and 80NSSC20K0389. E.A.C. acknowledges financial support from the IDEAS Fellowship, a research traineeship program funded by the National Science Foundation under grant DGE-1450006.
Partial support to this work was provided by 
the European Union Horizon 2020 Programme under the AHEAD2020 project (grant agreement number 871158). G.R. acknowledges support from the University of Maryland through the Joint Space Science Institute Prize Postdoctoral Fellowship.
The work of E.A.C., C.L.F., R.T.W., C.J.F and O.K. was supported by the US Department of Energy
through the Los Alamos National Laboratory, which is operated
by Triad National Security, LLC, for the National Nuclear Security
Administration of US Department of Energy (Contract No. 89233218CNA000001).

This work made use of data supplied by the UK \textit{Swift} Science Data Centre at the University of Leicester. The scientific results reported in this article are based on observations made by the Chandra X-ray Observatory. This research has made use of software provided by the Chandra X-ray Center (CXC) in the application package CIAO. Based on observations obtained at the international Gemini Observatory, a program of NSF's OIR Lab, which is managed by the Association of Universities for Research in Astronomy (AURA) under a cooperative agreement with the National Science Foundation on behalf of the Gemini Observatory partnership: the National Science Foundation (United States), National Research Council (Canada), Agencia Nacional de Investigaci\'{o}n y Desarrollo (Chile), Ministerio de Ciencia, Tecnolog\'{i}a e Innovaci\'{o}n (Argentina), Minist\'{e}rio da Ci\^{e}ncia, Tecnologia, Inova\c{c}\~{o}es e Comunica\c{c}\~{o}es (Brazil), and Korea Astronomy and Space Science Institute (Republic of Korea). The \textit{HST} data used in this work was obtained from the Mikulski Archive for Space Telescopes (MAST). STScI is operated by the Association of Universities for Research in Astronomy, Inc., under NASA contract NAS5-26555. Support for MAST for non-\textit{HST} data is provided by the NASA Office of Space Science via grant NNX09AF08G and by other grants and contracts. These results also made use of Lowell Observatory's Lowell Discovery Telescope (LDT), formerly the Discovery Channel Telescope. Lowell operates the LDT in partnership with Boston University, Northern Arizona University, the University of Maryland, and the University of Toledo. Partial support of the LDT was provided by Discovery Communications. LMI was built by Lowell Observatory using funds from the National Science Foundation (AST-1005313). We additionally made use of Astropy, a community-developed core Python package for Astronomy \citep{Astropy2018}. The afterglow modeling was performed in part on the George Washington University (GWU) Pegasus computer cluster and on the YORP cluster administered by the Center for Theory and Computation, which is part
of the Department of Astronomy at the University of Maryland (UMD).

\section*{Data Availability}

The data underlying this article will be shared on reasonable request to the corresponding author.




\bibliographystyle{mnras}
\bibliography{example} 



\appendix
\section{Supplementary Materials}

\begingroup
\renewcommand{\arraystretch}{1.5}
\begin{table*}
         \caption{Fit results of our afterglow modeling for GRB 200522A. We consider four model scenarios to describe the broadband transient: (i) a tophat jet model with standard forward shock emission and intrinsic extinction from the host galaxy (FS+Ext), (ii) forward shock emission with the addition of a simple blackbody component (FS+BB), (iii) forward shock emission and a blackbody component with intrinsic extinction (FS+BB+Ext), and (iv) a Gaussian jet model with forward shock emission and intrinsic extinction.
         This table represents the results of the posterior distribution for each fit, and in Table \ref{tab: fit_results_BB} we demonstrate the posterior distributions corresponding to the multiple modes allowed by each fit (see \S \ref{sec: 200522A_afterglow_FS} and \ref{sec: 200522A_afterglow_BB}).
         The median and 68\% confidence interval of the marginalized posterior distribution for each parameter is shown in rows 3-14. Row 15 shows the minimum $\chi^2$ value for each model. Rows 16-17 display the WAIC score of the expected log predictive density (\emph{elpd}), and the WAIC score difference, $\Delta$WAIC$_{\mathrm{elpd}}$, compared to the FS+Ext model. 
         }
        \begin{tabular}{lrrrr}
        \hline
        \hline
        \multirow{2}{*}{Parameter} & 
        \multicolumn{3}{c}{Tophat} & \multicolumn{1}{c}{Gaussian} \\
        \cmidrule(lr){2-4}  \cmidrule(lr){5-5}
      & FS + Ext & FS + BB & FS + BB + Ext & FS + Ext \\
         \hline
  $\log_{10} E_0$ (erg)       
 &    $50.69^{+0.08}_{-0.07}$
  &  $51.9^{+1.1}_{-0.7}$
  &  $52.0^{+1.1}_{-0.8}$
  & $52.1^{+1.3}_{-1.4}$
\\
  $\theta_c$ (rad)      
  & $0.16^{+0.04}_{-0.03}$
  & $0.22^{+0.38}_{-0.15}$
  & $0.19^{+0.39}_{-0.13}$
  & $0.26^{+0.40}_{-0.19}$
\\
 $\theta_v$ (rad)      
  & --
  & --
  & --
  & $0.30^{+0.31}_{-0.25}$
\\
 $\theta_w$ (rad)      
  & --
  & --
  & --
  & $0.71^{+0.55}_{-0.49}$
\\
  $\log_{10} E_j$  (erg) 
  & $48.8^{+0.2}_{-0.1}$
  &  $50.2^{+1.3}_{-0.9}$
  & $50.1^{+1.3}_{-0.8}$
  & $50.1^{+2.4}_{-1.0}$
\\
  $\log_{10} n_0$  (cm$^{-3}$) 
  &  $-2.7^{+0.3}_{-0.2}$
  & $-4.4^{+1.5}_{-1.1}$
  &   $-4.4^{+1.5}_{-1.2}$
  & $-3.6^{+1.0}_{-2.1}$
 \\
  $p$                   
  &    $2.32^{+0.19}_{-0.10}$
  & $2.44^{+0.16}_{-0.28}$
  & $2.41^{+0.18}_{-0.28}$
  & $2.74^{+0.26}_{-0.43}$
\\
  $\log_{10}\varepsilon_e$            
  &  $-0.52^{+0.03}_{-0.05}$
  & $-0.66^{+0.14}_{-0.21}$
  & $-0.66^{+0.13}_{-0.12}$
  & $-0.56^{+0.06}_{-0.16}$
\\
  $\log_{10}\varepsilon_B$            
  & $-0.57^{+0.07}_{-0.13}$
  &   $-2.0^{+1.0}_{-1.5}$
  & $-2.1^{+1.1}_{-1.5}$
  & $-1.1^{+0.5}_{-1.4}$
 \\
      \hline
       $E(B-V)$ (mag)
  & $0.16^{+0.08}_{-0.07}$
  & --
  &  $0.19^{+0.22}_{-0.14}$
  & $0.13^{+0.11}_{-0.07}$
 \\
 \hline
  $R$  ($10^{15}$ cm) 
  &  --
  & $2.2^{+1.1}_{-0.7}$
  &  $1.6^{+0.7}_{-0.4}$
  & --
 \\
     $T$  (K) 
  & --
  & $4100^{+900}_{-700}$
  &  $5800\pm1500$
  & --
\\
  \hline
  \hline
  $\chi^2$
  & 10.5
  & 3.8
  & 4.1
  & 5.4
  \\
  \hline
  WAIC$_{\mathrm{elpd}}$      
  &  $150\pm19$            
  & $151\pm19$
  &  $154\pm18$
  & $88\pm19$
 \\
  $\Delta$WAIC$_{\mathrm{elpd}}$
  &  --             
  & $1.2\pm3.2$
  & $4.2\pm3.5$
  & $-62\pm40$
 \\ 
  \hline
    \end{tabular}\\ 
   \label{tab: fit_results}
\end{table*}
\endgroup

\begin{figure*}
\centering
\includegraphics[width=2\columnwidth]{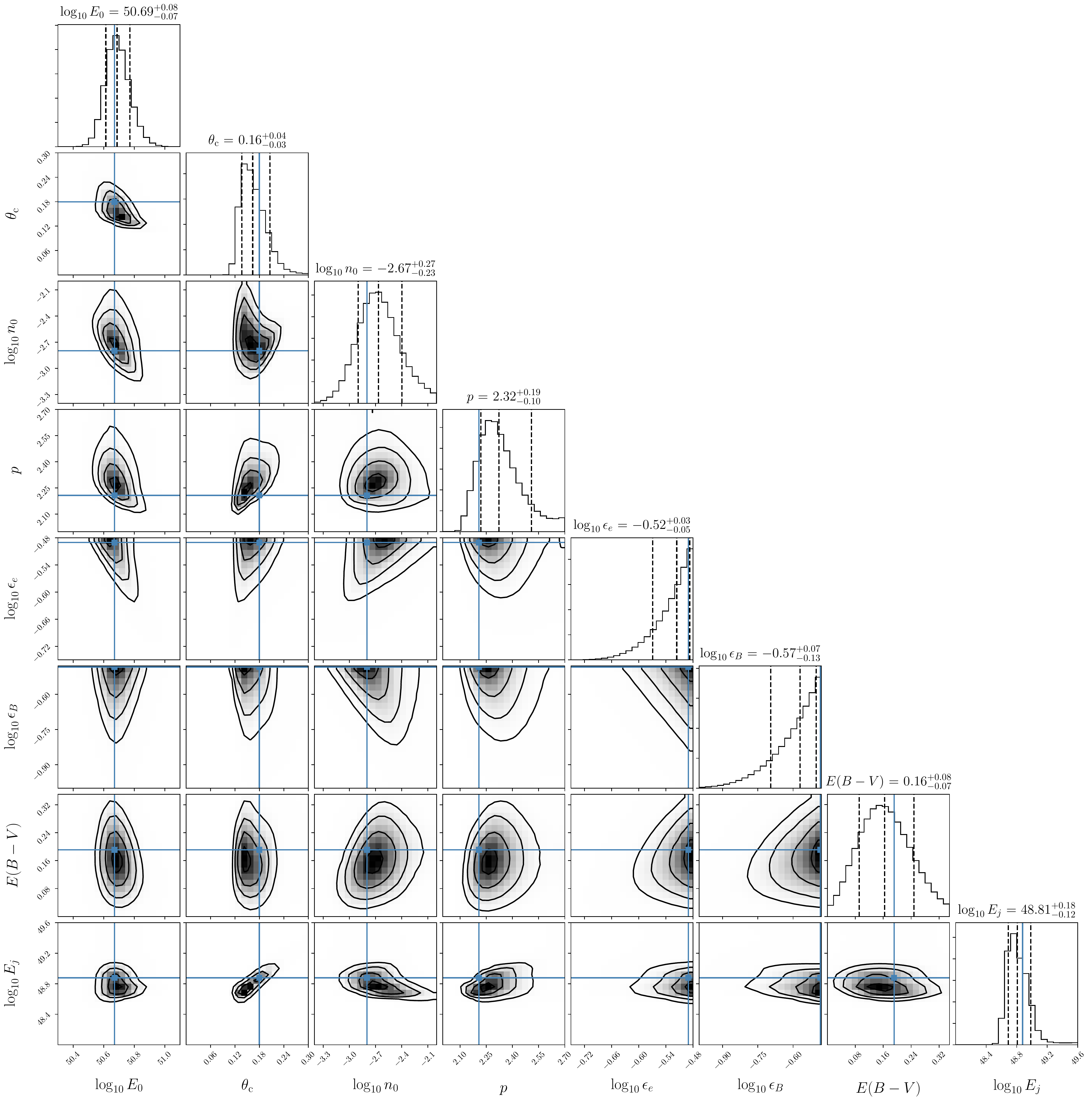}
\caption{MCMC fitting results for the FS+Ext model. The posterior probability distribution for the fit parameters from our MCMC fitting routine are shown in one-dimension (diagonal) and two-dimensions (off-diagonal). Values corresponding to the maximum posterior probability (best fit) are marked by blue lines. The contours (off-diagonal) and dashed lines (diagonal) correspond to the 16\%, 50\%, and 84\% quantiles of the marginalized posterior probability for each parameter. 
}
\label{fig: 200522A_fs+ext_corner}
\end{figure*}

\begin{figure*}
\centering
\includegraphics[width=2\columnwidth]{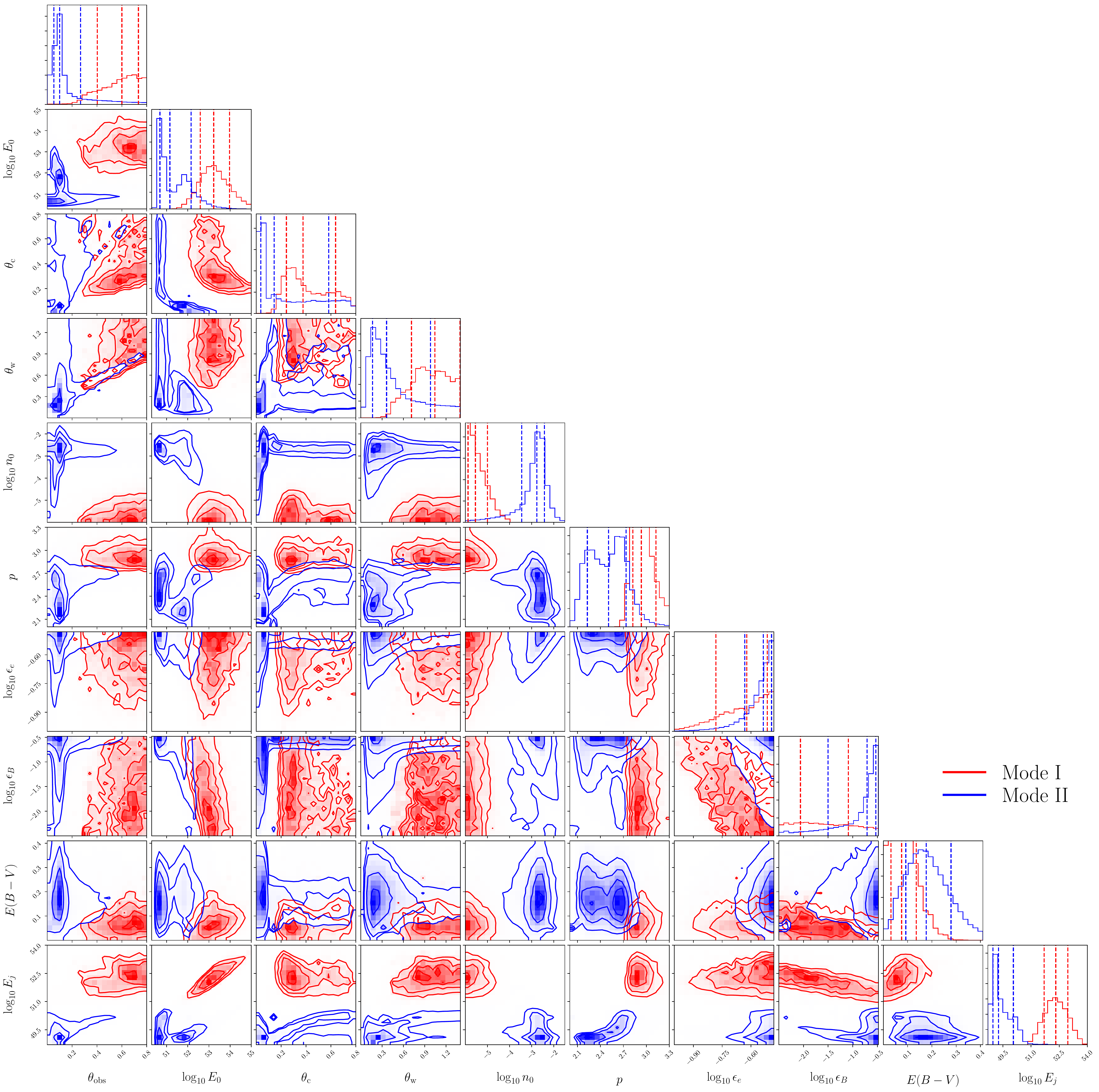}
\caption{
Same as Figure \ref{fig: 200522A_fs+ext_corner} for the Gauss+FS+Ext model corresponding to a Gaussian structured jet profile. 
The two modes (see Table \ref{tab: fit_results_BB}) of the fit are shown in \emph{red} and \emph{blue} for \emph{Mode I} and \emph{Mode II}, respectively. Each mode is preferred with the same weight by the MCMC fit, i.e., $\sim50\%$ of walkers prefer each mode.
}
\label{fig: 200522A_gauss+fs+ext_corner}
\end{figure*}

\begin{figure*}
\centering
\includegraphics[width=2\columnwidth]{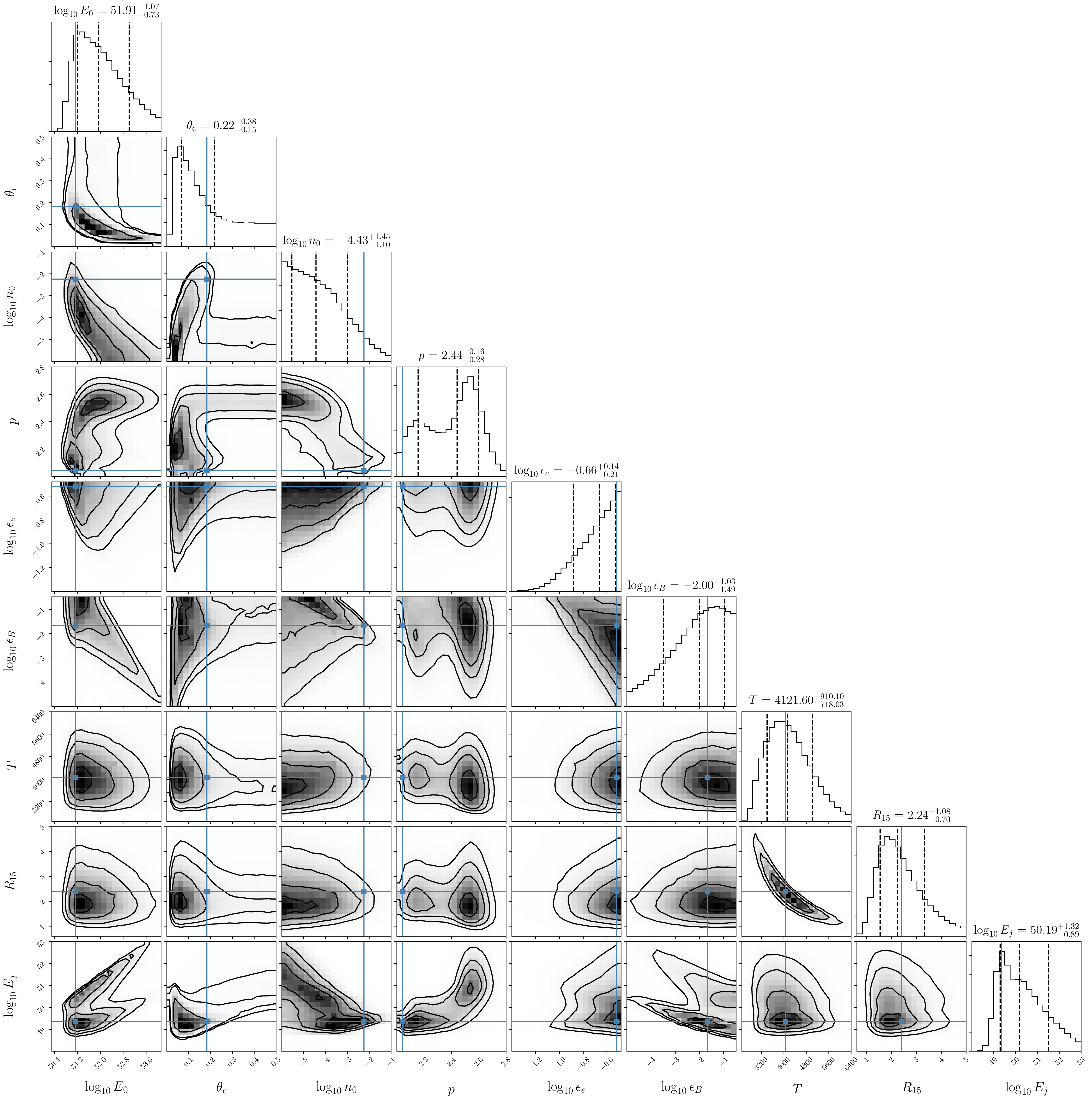}
\caption{
Same as Figure \ref{fig: 200522A_fs+ext_corner} for the FS+BB model. 
$R_{15}$ denotes the emission radius of the blackbody in units of $10^{15}$ cm, i.e., $R=R_{15}10^{15}$ cm.
}
\label{fig: 200522A_fs+bb_corner}
\end{figure*}

\begin{figure*}
\centering
\includegraphics[width=2\columnwidth]{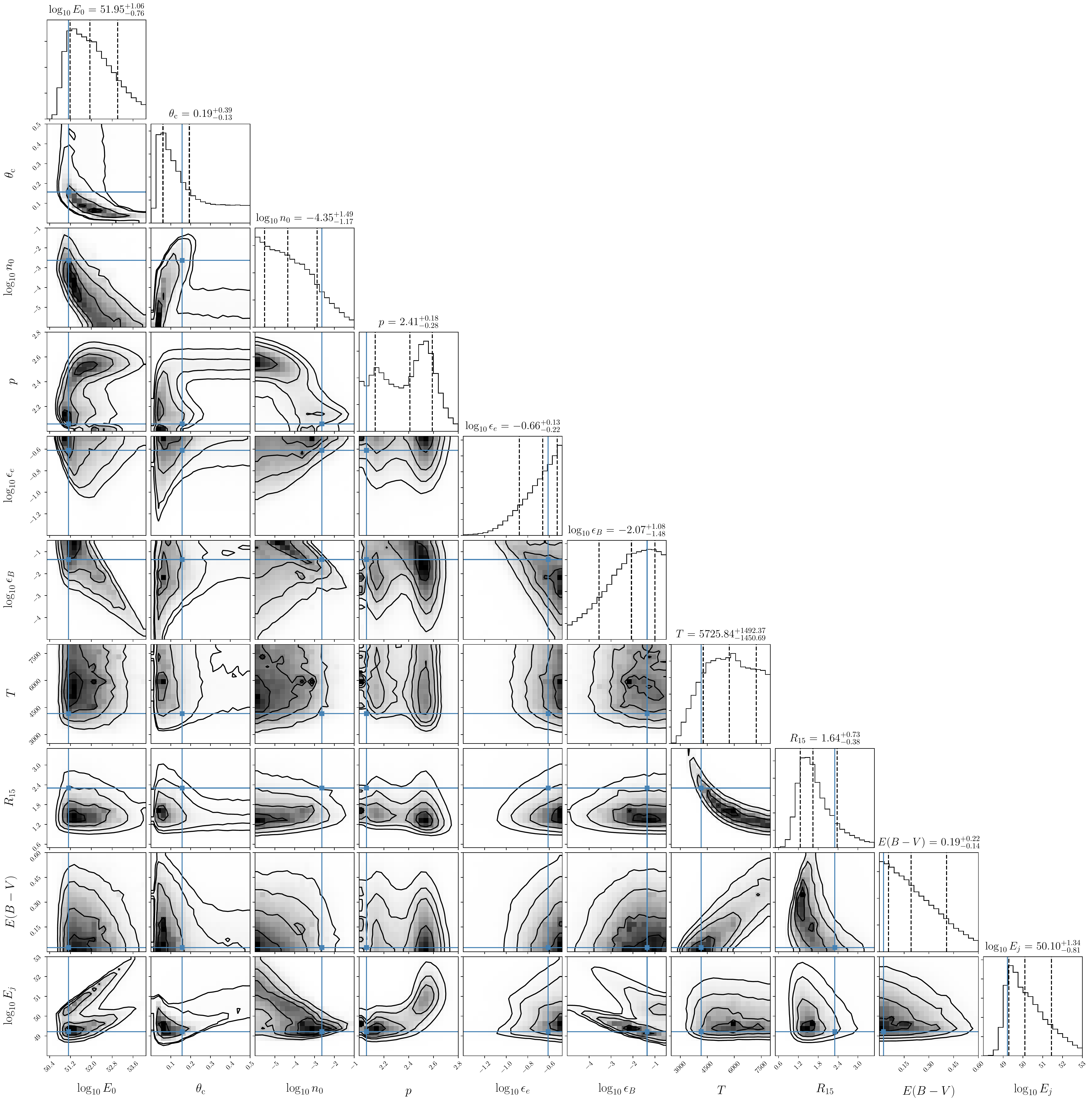}
\caption{Same as Figure \ref{fig: 200522A_fs+ext_corner} for the FS+BB+Ext model. 
}
\label{fig: 200522A_fs+bb_corner2}
\end{figure*}



\bsp	
\label{lastpage}
\end{document}